\documentclass[twocolumn]{aastex631}

\usepackage{footnote}

\shorttitle{Coronal Magnetic Field Measurements}
\shortauthors{Chen et al.}

\newcommand{\sect}[1]{Section\,\ref{#1}}
\newcommand{\sects}[1]{Sections\,\ref{#1}}
\newcommand{\figa}[1]{Figure\,~\ref{#1}}

\newcommand{\tab}[1]{Table\,\ref{#1}}

\newcommand{\eqn}[1]{Equation\,(\ref{#1})}
\newcommand{\eqns}[1]{Equations\,(\ref{#1})}

\definecolor{orange}{rgb}{1,0.4,0.}
\graphicspath{{./}{figures/}}

\begin{document}

\title{Forward Modeling of Solar Coronal Magnetic Field Measurements Based on a Magnetic-field-induced Transition in Fe~{\sc{x}}}

\correspondingauthor{Wenxian Li, Hui Tian}
\email{wxli@nao.cas.cn, huitian@pku.edu.cn}

\author{Yajie Chen}
\affiliation{School of Earth and Space Sciences, Peking University, 100871 Beijing, China}

\author{Wenxian Li}
\affiliation{Key Laboratory of Solar Activity, National Astronomical Observatories, Chinese Academy of Sciences, Beijing 100012, China}

\author{Hui Tian}
\affiliation{School of Earth and Space Sciences, Peking University, 100871 Beijing, China}
\affiliation{Key Laboratory of Solar Activity, National Astronomical Observatories, Chinese Academy of Sciences, Beijing 100012, China}

\author{Feng Chen}
\affiliation{School of Astronomy and Space Science, Nanjing University, Nanjing 210023, China}
\affiliation{Key Laboratory for Modern Astronomy and Astrophysics (Nanjing University), Ministry of Education, Nanjing 210023, China}
\affiliation{Key Laboratory of Solar Activity, National Astronomical Observatories, Chinese Academy of Sciences, Beijing 100012, China}

\author{Xianyong Bai}
\affiliation{Key Laboratory of Solar Activity, National Astronomical Observatories, Chinese Academy of Sciences, Beijing 100012, China}
\affiliation{School of Astronomy and Space Science, University of Chinese Academy of Sciences, Beijing 100049, China}

\author{Yang Yang}
\affiliation{Shanghai EBIT laboratory, Institute of Modern physics, Fudan University, Shanghai, China}
\affiliation{Key Laboratory of Nuclear Physics and Ion-beam Application (MOE), Fudan University, Shanghai 200433, China}

\author{Zihao Yang}
\affiliation{School of Earth and Space Sciences, Peking University, 100871 Beijing, China}

\author{Xianyu Liu}
\affiliation{School of Earth and Space Sciences, Peking University, 100871 Beijing, China}

\author{Yuanyong Deng}
\affiliation{Key Laboratory of Solar Activity, National Astronomical Observatories, Chinese Academy of Sciences, Beijing 100012, China}
\affiliation{School of Astronomy and Space Science, University of Chinese Academy of Sciences, Beijing 100049, China}

\begin{abstract}
It was recently proposed that the intensity ratios of several extreme ultraviolet spectral lines from the Fe~{\sc{x}} ion can be used to measure the solar coronal magnetic field based on the magnetic-field-induced transition (MIT) theory. 
To verify the suitability of this method, we performed forward modeling with a three-dimensional radiation magnetohydrodynamic model of a solar active region. 
Intensities of several spectral lines from Fe~{\sc{x}} were synthesized from the model. Based on the MIT theory, intensity ratios of the MIT line Fe~{\sc{x}} 257 {\AA}~to several other Fe~{\sc{x}}~lines were used to derive the magnetic field strengths, which were then compared with the field strengths in the  model.
We also developed a new method to simultaneously estimate the coronal density and temperature from the Fe~{\sc{x}} 174/175 and 184/345 {\AA} line ratios. Using these estimates, we demonstrated that the MIT technique can provide reasonably accurate measurements of the coronal magnetic field in both on-disk and off-limb solar observations. 
Our investigation suggests that a spectrometer that can simultaneously observe the Fe~{\sc{x}} 174, 175, 184, 257, and 345 {\AA} lines and allow an accurate radiometric calibration for these lines is highly desired to achieve reliable measurements of the coronal magnetic field. We have also evaluated the impact of the uncertainty in the Fe~{\sc{x}} 3p$^4$ 3d $^4$D$_{5/2}$ and $^4$D$_{7/2}$ energy difference on the magnetic field measurements. 

%
%
%
%

\end{abstract}


\section{Introduction} \label{sec:intro}
The magnetic field couples different layers of the solar atmosphere and governs most types of solar activity. 
Therefore, routine and accurate measurements of the magnetic field at all atmospheric layers are of critical importance.
However, such measurements have only been achieved at the photospheric level. More than one century has passed since the first measurement of the solar magnetic field, we still have very limited knowledge of the magnetic field above the photosphere, especially in the corona \citep[e.g.,][]{Wiegelmann2014}. 

Extrapolation from the observed photospheric magnetograms is a useful tool to construct the magnetic field structures above the photosphere \citep[e.g.,][]{Schatten1969,Aly1984,Chifu2017,Aschwanden2013,Zhu2018,Wiegelmann2021}.
In addition, with the forward modeling approach, a combination of magnetohydrodynamic (MHD) or magnetic field models and infrared or extreme ultraviolet (EUV) observations could also help us understand coronal magnetic field structures \citep[e.g.,][]{Liu2008,Dove2011,2013ApJ...770L..28B,Rachmeler2013,Gibson2016,Jibben2016,Li2017,Chen2018,Zhao2019,Zhao2021}.
%
Although widely used for investigations of coronal magnetism, these models are highly dependent on various assumptions, which are not always valid in the solar atmosphere.
Thus, it is imperative to conduct more direct measurements of the coronal magnetic field.

Zeeman-effect-based methods have been well developed to measure the magnetic field in the photosphere \citep[e.g., ][]{Iniesta2016}.
However, coronal magnetic field measurements using the Zeeman effect are more challenging because of the negligible Zeeman splitting and weak spectropolarimetric signatures.
\citet{Lin2000,Lin2004} have attempted to measure the coronal magnetic field in off-limb active regions through the Zeeman effect using the strong near-infrared Fe~{\sc{xiii}} 10747 {\AA} line.  
However, in order to obtain a high enough signal-to-noise ratio, they integrated the data over $\sim$70 minutes. 
Thus, this technique is difficult to measure short-term variations of the coronal magnetic field.

Radio imaging observations have also been used for coronal magnetic field measurements in both quiet active regions \citep[e.g.,][]{Akhmedov1982,Ryabov1999,Miyawaki2016,Anfinogentov2019} and flaring/eruptive structures \citep[e.g.,][]{Raja2014,Tan2016,Gary2018,Fleishman2020,Chen2020,Yu2020,Zhu2021}.
These measurements, however, rely on the determination of radio emission mechanisms from observations, which is not always straightforward.

Observations of some types of coronal oscillations and waves could be used to infer physical parameters of the coronal plasma such as the magnetic field strength through magnetoseismology \citep{Nakariakov2020}.
Earlier studies based on impulsively excited decaying oscillations could only provide an average value or 1D distribution of the magnetic field strength of the oscillating coronal structures \citep[e.g.,][]{Nakariakov2001,Chen2011,Chen2015,Li2018,Su2018}. However, if we apply the magnetoseismology technique to the more ubiquitous waves and oscillations in the corona \citep[e.g., ][]{Jess2016,Tomczyk2007,Tian2012,Wang2012,Morton2015}, 2D distributions or temporal evolution of the coronal magnetic field strength could be derived \citep[][]{Long2017,Magyar2018}. 
More recently, \citet{Yang2020b,Yang2020a} mapped the global coronal magnetic field for the first time by applying magnetoseismology to the pervasive propagating transverse waves in the corona. 
However, their technique can only measure the plane-of-sky component of the magnetic field and will fail in regions where coronal mass ejections occur.

Recently, a new diagnostic technique was proposed to measure the coronal magnetic field by using the magnetic-field-induced transition (MIT) of the Fe~{\sc{x}} 257 {\AA} line, which is based on a quantum interference effect induced by an accidental and pseudo-degeneracy of energy levels, 3p$^4$3d~$^4$D$_{5/2}$ and $^4$D$_{7/2}$, of Fe~{\sc{x}} in the presence of an external magnetic field \citep{Li2015, Li2016}.
The Fe~{\sc{x}} ion undoubtedly exists in the corona, and the EUV Fe~{\sc{x}} solar spectra, including the 257 {\AA} line, have been routinely observed by the EUV Imaging Spectrometer \citep[EIS; ][]{Culhane2007} onboard the Hinode satellite \citep{Kosugi2007}. 
Hence, this new technique might allow routine measurements of the coronal magnetic field both on the solar disk and in off-limb regions.
\citet{Si2020} first applied this technique to Hinode/EIS observations of an active region and obtained a magnetic field strength of $\sim$270 G.
\citet{Landi2020b} further developed this technique and obtained two-dimensional magnetic field maps for several active regions in both on-disk and off-limb observations with Hinode/EIS. 
The same technique was also used by \citet{Brooks2021} to determine the magnetic field strength in the source region of a solar energetic particle event, which was found to be in the range of 325 to 550 G. 
More recently, \citet{Landi2021} applied this technique to EIS observations of a flare and found notable changes of the total magnetic field strength during the flare process.
All these investigations demonstrated the potential of the MIT technique for measurements of the coronal magnetic field in solar active regions, although the results are associated with large uncertainties that are mainly related to uncertainties in the radiometric calibration of Hinode/EIS and the atomic data.

However, we realize that the MIT technique has not yet been validated. So here we aim to verify the suitability and evaluate the accuracy of this technique in coronal magnetic field measurements through the approach of forward modeling. 
We introduce our 3D radiation MHD model and spectral line synthesis from the model in \sect{sec:model}, compare the coronal magnetic field strength derived from the MIT technique to the value in the model in \sects{sec:inver_uni} and \ref{sec:inver_lsq}, and investigate the impact of the uncertainty in the energy separation between the $^4$D$_{5/2}$ and $^4$D$_{7/2}$ levels on the magnetic field estimation in \sect{sec:dE}.
Finally, we summarize our results and point out future directions to achieve reliable measurements of the coronal magnetic field through the MIT technique in \sect{sec:sum}.

\section{Radiation MHD model and emission line synthesis} \label{sec:model}

We first created an atomic database for several spectral lines from the Fe~{\sc{x}} ion, which is described in \sect{subsec:atom_model}. 
Then we took a snapshot of a 3D radiation MHD model of an active region (\sect{subsec:solar_model}) and synthesized emission of the Fe~{\sc{x}} lines of interest (\sect{subsec:emission}). 
%

\subsection{Fe~{\sc{x}} lines used for magnetic field diagnostics}   \label{subsec:atom_model}

\begin{figure*} 
\centering {\includegraphics[width=14cm]{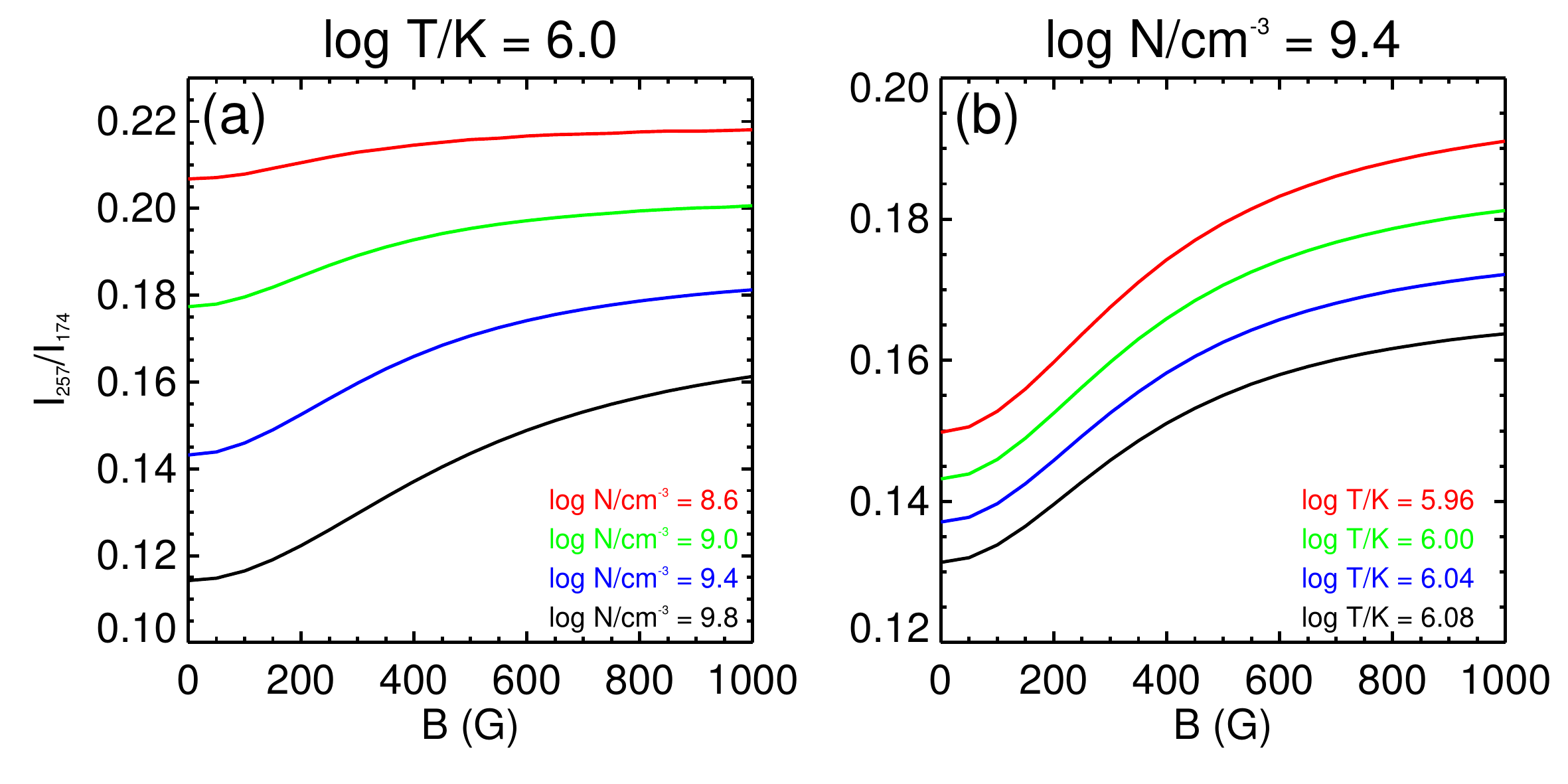}} 
\caption{Intensity ratio of the Fe~{\sc{x}} 257 and 174 {\AA} lines, $I_{257}/I_{174}$, as a function of the magnetic field strength at a temperature of 10$^{6.00}$ K and densities of 10$^{8.6}$, 10$^{9.0}$, 10$^{9.4}$, and 10$^{9.8}$ cm$^{-3}$ (a),  and at a density of 10$^{9.4}$ cm$^{-3}$ and temperatures of 10$^{5.96}$, 10$^{6.00}$, 10$^{6.04}$, and 10$^{6.08}$ K (b). } \label{f1}
\end{figure*}

For a detailed description of the MIT process in Fe~{\sc{x}}, we refer to \citet{Li2015,Li2016}. Here we just present a brief introduction. 
An electric dipole (E1) transition can occur from the 3p$^4$3d~$^4$D$_{5/2}$ level to the ground state 3p$^5$~$^2$P$_{3/2}$.
In the absence of an external magnetic field, the 3p$^4$3d~$^4$D$_{7/2}$ level can only decay to the ground state through a forbidden magnetic quadruple (M2) transition.
In the presence of an external field, through a quantum mixing of the $^4$D$_{7/2}$ level with the close-by $^4$D$_{5/2}$ level, an MIT decay channel from the $^4$D$_{7/2}$ level to the ground state can be induced, {with a transition rate increasing with the magnetic field strength}. 
The transition probability of MIT, $A_{\mathrm{MIT}}$, according to the first-order perturbation theory, depends on both the magnetic field strength (\textit{B}) and the energy separation ($\Delta E$) between the $^4$D$_{7/2}$ and $^4$D$_{5/2}$ levels:
\begin{equation}
    A_{\mathrm{MIT}} \propto \frac{B^2}{{(\Delta E)}^2}
    \label{eq:amit}
\end{equation}
It is worth noting that the wavelength difference between the E1 (257.259 {\AA}) and M2/MIT (257.261 {\AA}) lines is too small to be resolved in observations. So in the following the 257 {\AA} line intensity refers to the total intensity of the E1, M2, and MIT lines.

The diagnostic technique is based on a simple concept: 
the intensity of the MIT emission and thereby the 257 {\AA} (MIT+M2+E1) line intensity is directly and significantly affected by the local magnetic field strength; it therefore can be used to measure the magnetic field strength by measuring the intensity ratio of a line pair, i.e., the 257 {\AA} line and a reference Fe~{\sc{x}} line that is not sensitive to the magnetic field. 
The Fe~{\sc{x}} 174, 175, 177, 184, and 255 {\AA} lines can be observed simultaneously along with the 257 {\AA} line by \textit{Hinode}/EIS, and therefore could be adopted as candidates of the reference line. 
The atomic transitions corresponding to these lines are listed in \tab{tab:lines}.
We used CHIANTI database \citep[version 10.0; ][]{1997A&AS..125..149D,2021ApJ...909...38D} to calculate the line emissivities.
The original source for these data are \citet{DelZanna2012} for level energies and collision data, \citet{Wang2020} for radiative transition data, and  \citet{Li2021} for $A_\mathrm{MIT}$ values.

\begin{table}
\caption{Fe~{\sc{x}} lines used in this study.}
\centering
\begin{tabular}{c|c}
\hline\hline
 Wavelength$^a$ ({\AA})    & Upper level $\to$ Lower level      \\ \hline
 174.531     & 3s$^2$ 3p$^4$ 3d $^2$D$_{5/2}$ $\to$  3s$^2$ 3p$^5$ $^2$P$_{3/2}$\\ \hline
 175.263    & 3s$^2$ 3p$^4$ 3d $^2$D$_{3/2}$ $\to$  3s$^2$ 3p$^5$ $^2$P$_{1/2}$ \\ \hline
 177.240    & 3s$^2$ 3p$^4$ 3d $^2$P$_{3/2}$ $\to$  3s$^2$ 3p$^5$ $^2$P$_{3/2}$  \\ \hline
 184.537    & 3s$^2$ 3p$^4$ 3d $^2$S$_{1/2}$ $\to$  3s$^2$ 3p$^5$ $^2$P$_{3/2}$ \\ \hline
 255.393    & 3s$^2$ 3p$^4$ 3d $^4$D$_{1/2}$ $\to$  3s$^2$ 3p$^5$ $^2$P$_{3/2}$ \\ \hline
 257.259    & 3s$^2$ 3p$^4$ 3d $^4$D$_{5/2}$ $\to$  3s$^2$ 3p$^5$ $^2$P$_{3/2}$ \\ \hline
 257.261    & 3s$^2$ 3p$^4$ 3d $^4$D$_{7/2}$ $\to$  3s$^2$ 3p$^5$ $^2$P$_{3/2}$ \\ \hline
 345.738    & 3s 3p$^6$ $^2$S$_{1/2}$        $\to$  3s$^2$ 3p$^5$ $^2$P$_{3/2}$ \\ \hline
\end{tabular} 
\\\footnotesize{$^a$ The wavelengths are obtained from the CHIANTI database.}\\
\label{tab:lines}
\end{table}

%
These line ratios also depend on the electron density and temperature. Taking the 257 and 174 {\AA} line pair as an example, we show the intensity ratio ($I_{257}/I_{174}$) as a function of the magnetic field at different densities and temperatures in \figa{f1}. 
\figa{f1}(a) shows the $I_{257}/I_{174}$ intensity ratio as a function of the magnetic field strength for four different values of the electron density, i.e., 10$^{8.6}$, 10$^{9.0}$, 10$^{9.4}$, and 10$^{9.8}$ cm$^{-3}$. These densities values are typical for different structures in the solar corona. A fixed temperature of 10$^{6.0}$ K, around which the contribution functions of the Fe~{\sc{x}} lines peak, was adopted.
%
%
As expected, since the upper $^4$D$_{7/2}$ level of the 257 {\AA} line is metastable, the $I_{257}/I_{174}$ ratio is also sensitive to electron density.
%
%
In \figa{f1}(b) we took a typical density of active regions (10$^{9.4}$~cm$^{-3}$) and four different temperatures of 10$^{5.96}$, 10$^{6.00}$, 10$^{6.04}$, and 10$^{6.08}$ K.
It is clear that the line ratio also changes with temperature. A similar behavior is also seen for the ratios of 257 {\AA} and other reference lines.
\figa{f1} indicates that we need to determine both the temperature and density before deriving the magnetic field strength using the MIT technique.
Fortunately, the Fe~{\sc{x}} 184/345 {\AA} line pair can be used to diagnose the electron temperature, and the Fe~{\sc{x}} 174/175 {\AA} line pair is good for coronal density diagnostics \citep{Phillips2008,Del_Zanna2018}. 

Another important atomic parameter that affects the accuracy of the coronal magnetic field measurement is the energy separation $\Delta E$ between the $^4$D$_{7/2}$ and $^4$D$_{5/2}$ levels (see \eqn{eq:amit}).
%
There have been several attempts to estimate the values of $\Delta E$ from solar observations as well as laboratory measurements \citep{1979ApJ...227L.107S,1994ApJS...91..461T,1998ApJS..119..255B,Li2016,Judge2016}.
More recently, based on observations with the Solar Ultraviolet Measurements of Emitted Radiation \citep[SUMER,][]{Wilhelm1995} onboard the Solar and Heliospheric Observatory (SOHO), \citet{Landi2020a} obtained a value of $\Delta E=2.29\pm0.50~$cm$^{-1}$, which was adopted for the computation of $A_\mathrm{MIT}$ and thereby the magnetic field diagnostics in the present work.
%
Most analyses in this study were performed using $\Delta E=2.29$~cm$^{-1}$. In \sect{sec:dE}, we adopted $1.79$ and $2.79$~cm$^{-1}$ as the lower and upper limits of $\Delta E$, respectively, to investigate the impact of the uncertainty in the energy separation on magnetic field measurements.

Finally, we created lookup tables of the contribution functions $G(T,n_e,B)$ for the Fe~{\sc{x}} lines listed in \tab{tab:lines} at a large number of electron temperatures (log$_{10}T$, from 5.6 to 6.4 with an interval of 0.002), electron densities (log$_{10}n_e$, from 6 to 15 with an interval of 0.008), and magnetic field strengths ($B$, from 0 to 6000 G with an interval of 5 G). 
Since these Fe~{\sc{x}} spectral lines are emitted from the same ion, it is reasonable to assume that they form in {roughly} the same layers/structures of the corona. 
In other words, these lines sample {almost} the same plasma in the line of sight (LOS). In such a case, the intensity ratio of two lines may be approximated by the ratio of their contribution functions at a certain density, temperature and magnetic field.
%

\subsection{Radiation MHD model of an active region} \label{subsec:solar_model}

\begin{figure*} 
\centering {\includegraphics[width=12cm]{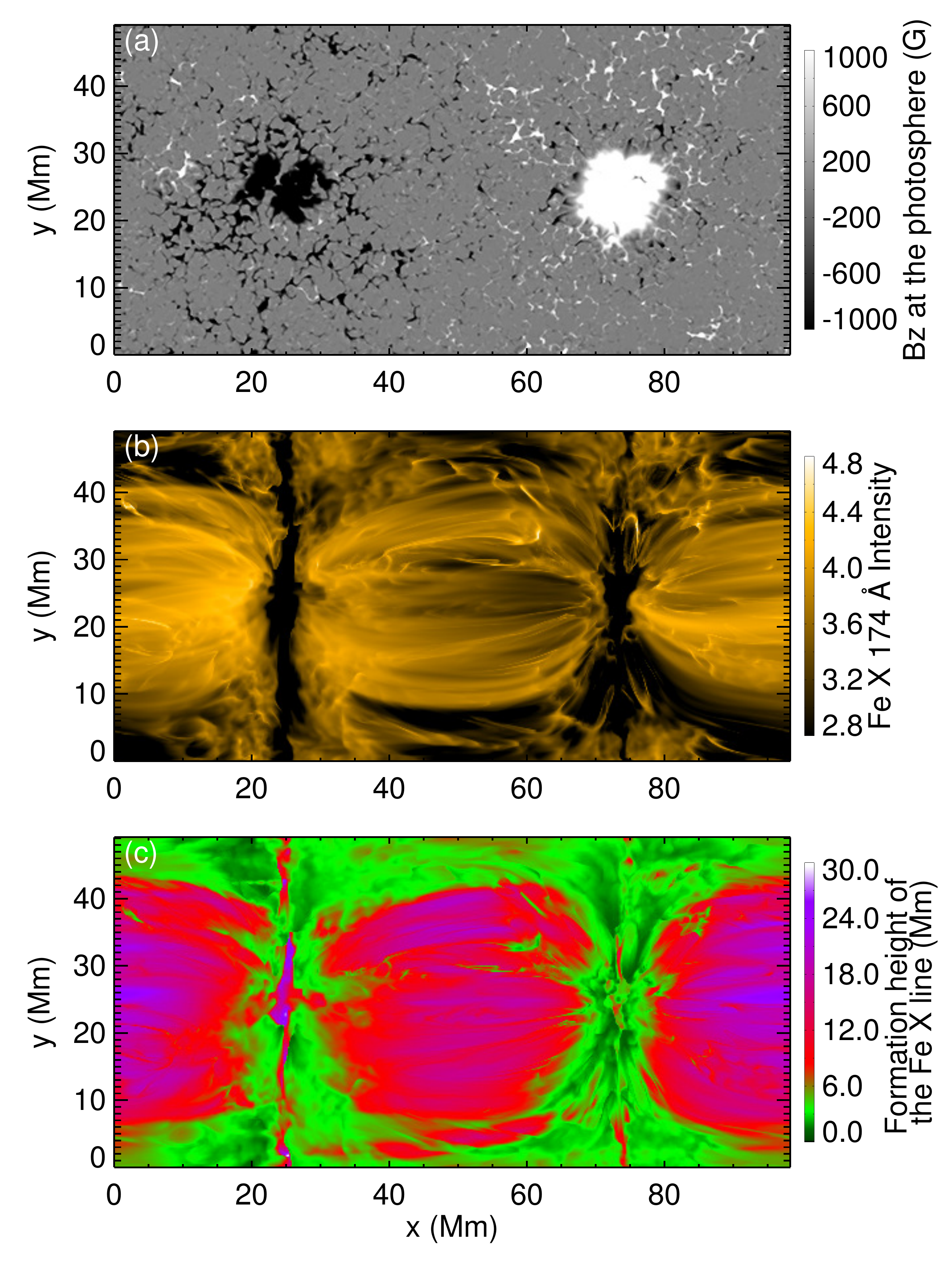}} 
\caption{Overview of the solar active region model. 
(a) Vertical component of the photospheric magnetic field. 
(b) Intensity map of the Fe~{\sc{x}} 174 {\AA} line when viewed along the vertical direction, shown in a logarithm scale and arbitrary unit.
(c) Map of the average formation height of the Fe~{\sc{x}} 174 {\AA} line.
} \label{f3}
\end{figure*}

\figa{f1} shows that the line ratio $I_{257}/I_{174}$ is more sensitive to the magnetic field when the density is higher, which implies that this technique is more suitable for magnetic field diagnostics in active regions where the density is usually higher.
%
%
So far, the MIT-based magnetic field measurements were all made for active regions \citep{Si2020,Landi2020b,Landi2021,Brooks2021}.
Therefore, we constructed a 3D MHD model of an active region to verify the suitability of this new technique and evaluate the accuracy of the measured magnetic field strength.

We used the MURaM radiation MHD code \citep{muram2005,Rempel2017} to simulate an active region. The model contains an area of 98.304$\times$49.152 Mm$^2$ in the horizontal direction and extends from 7.488 Mm below to 41.6 Mm above the photosphere in the vertical direction. The horizontal and vertical grid spacings in the model are 192 and 64 km, respectively.
The lateral boundary are periodic. The top boundary is half open and only allows for outflows. A torus flux tube imported from the bottom boundary \citep{Rempel2014} emerges into the photosphere and gives to the active region.

When the active region evolves into a stable state, we took a snapshot from the model. 
\figa{f3}(a) shows the vertical component of the photospheric magnetic field in the snapshot. 
There is a bipolar sunspot pair with magnetic field strengths of several thousand Gauss in the simulation box.
The plasma temperatures at coronal heights are of the order of $\sim$10$^{5.9-6.3}$ K, which are typical for the solar corona. Under the situation of ionization equilibrium, the Fe~{\sc{x}} lines are known to form in this temperature range.
%

\subsection{Synthetic coronal emissions}
\label{subsec:emission}

In order to perform forward modeling, we synthesized emissions of the Fe~{\sc{x}} lines listed in \tab{tab:lines}.
The emissivity $\epsilon_i$ of the Fe~{\sc{x}} lines can be written as
\begin{equation}
    \epsilon_i=G_i(T,n_e,B)n_e^2  \label{eq:em}
\end{equation}
where $G_i(T,n_e,B)$ is the contribution function and the subscript $i$ ($i=$174, 175, 177, 184, 255, 257, and 345) corresponds to the wavelength of an Fe~{\sc{x}} line in the unit of angstrom.
We first calculated the emissivity of each line at each grid point and then integrated the emissivity along the LOS. Hence, the line intensity can be written as
\begin{equation}
    I_i=\int_{LOS} \epsilon_i ds  \label{eq:int}
\end{equation}
\figa{f3}(b) shows the Fe~{\sc{x}} 174 {\AA} line intensity map after integration of the emissivity along the vertical direction, assuming location of the active region at the solar disk center.
Typical coronal loop structures are clearly present in the synthesized coronal image. Similar structures are also seen in the intensity maps of other Fe X lines.

Furthermore, we defined an emission-weighted formation height of the Fe~{\sc{x}} 174 {\AA} line as
\begin{equation}
    h_{em}=\frac{\int_z \epsilon_{174}(z) \cdot z dz}{\int_z \epsilon_{174}(z) dz}  \label{eq:em_h}
\end{equation}
where $\epsilon$ and $z$ are the emissivity and height at each grid point, respectively.
It is worth noting that the emission can extend over a wide range of height, and thus the formation height obtained from \eqn{eq:em_h} can only be regarded as an average height of the Fe~{\sc{x}} 174 {\AA}--emitting plasma. 
A map of the formation height of the Fe~{\sc{x}} 174 {\AA} line is shown in \figa{f3}(c), from which we can see that $h_{em}$ varies at different regions and extends from below 6 Mm near the sunspots (at lower parts of the coronal loops) to $10-20$ Mm in higher parts of the loops. Similar results were also obtained for other Fe~{\sc{x}} lines.
%


%

\section{MIT-based magnetic field measurements using a uniform temperature distribution} \label{sec:inver_uni}

\subsection{Density diagnostics and magnetic field measurements} \label{subsec:diag_Trho_uni}
%
As discussed in \sect{subsec:atom_model}, the intensity ratios of the Fe~{\sc{x}} line pairs used to derive the magnetic field strength are also temperature- and density-sensitive.
Thus, we need to determine the temperature and density before conducting magnetic field measurements.
%
%
%
%
We first adopted a fixed temperature of $T_0=10^{6.0}$ K, around which the contribution functions of the Fe~{\sc{x}} lines reach their peaks;
%
then we used the 174/175 {\AA} line ratio for the density estimation.
%
%
Since the intensity ratio $I_{174}/I_{175}$ is barely sensitive to the magnetic field, we can determine the electron density using the ratio of the contribution functions of the Fe~{\sc{x}} 174 and 175 {\AA} lines calculated in \sect{subsec:atom_model}, as
\begin{equation}
    \frac{I_{174}}{I_{175}}=\frac{n_e^2\cdot G_{174}(T_0,n_e,B)}{n_e^2\cdot G_{175}(T_0,n_e,B)}\approx \frac{G_{174}(T_0,n_e,B=0)}{G_{175}(T_0,n_e,B=0)} \label{r_174_175}
\end{equation}

Once the density at each pixel (e.g., each [x,y] position in \figa{f3}) in the coronal image, $n_e^*$, is determined, the intensity ratio between the Fe~{\sc{x}} 257 {\AA} line and a reference line ${I_{257}}/{I_{i}}$ ($i=174,175,177,184,255$) only depends on the magnetic field strength,
\begin{equation}
    \frac{I_{257}}{I_{i}}=\frac{G_{257}(T_0,n_e^*,B)}{G_{i}(T_0,n_e^*,B)} \label{r_257_i}
\end{equation}
meaning that we can derive the coronal magnetic field strength from the ${I_{257}}/{I_{i}}$ value at each pixel. 

%

\begin{figure*} 
\centering {\includegraphics[width=\textwidth]{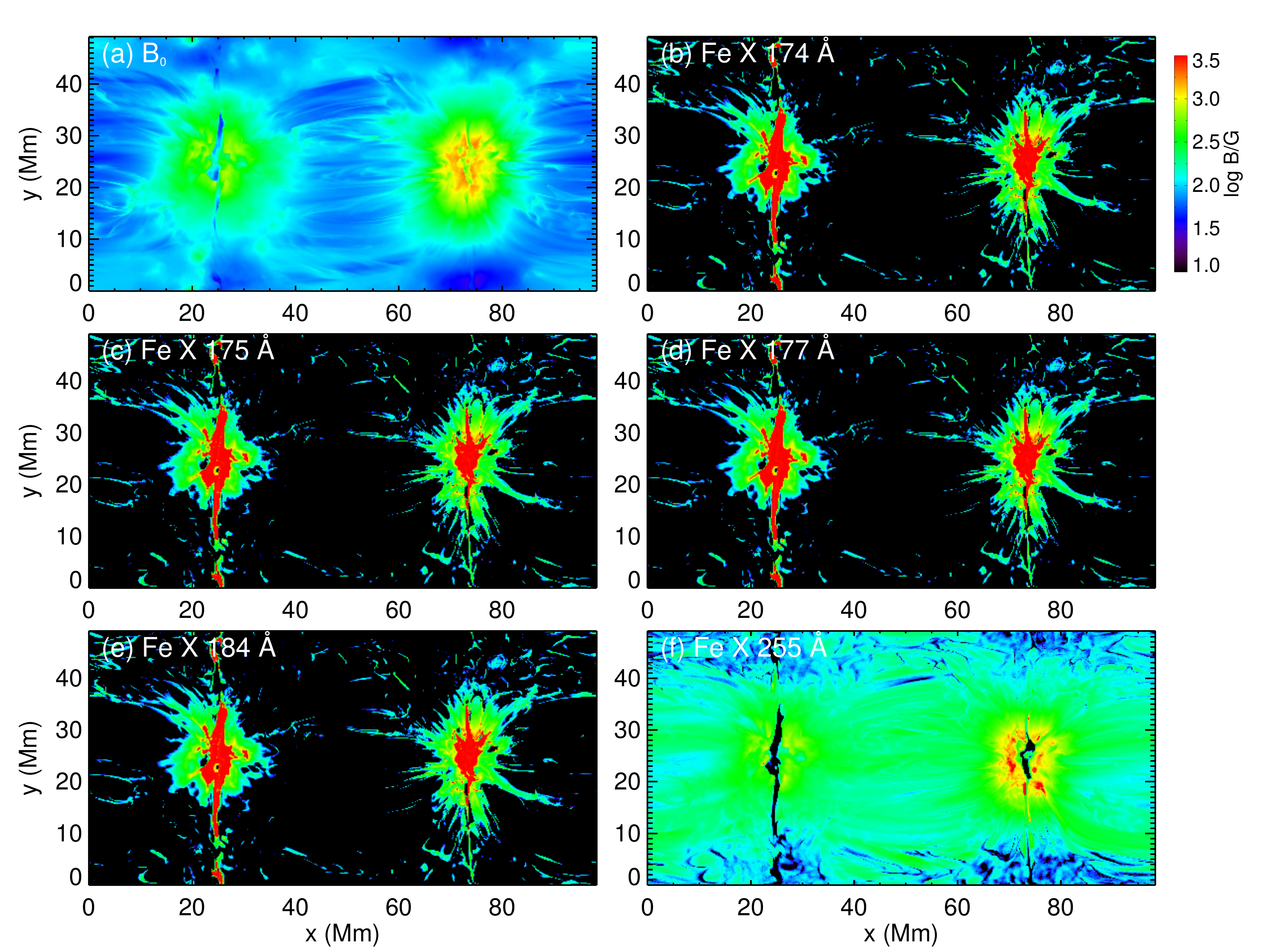}} 
\caption{(a) Coronal magnetic field strength in the model calculated using \eqn{eq:em_b}. (b--f) The derived magnetic field strength using the intensity ratios of the Fe~{\sc{x}} 257/174, 257/175, 257/177, 257/184, and 257/255 {\AA} line pairs, respectively, based on the MIT technique. We assumed a fixed temperature of 10$^6$ K at all pixels for the diagnostics of electron density and magnetic field. 
} \label{f9}
\end{figure*}

\subsection{Coronal magnetic field in the MHD model} \label{subsec:b0}

%
Because all the spectral lines we used in this work are emitted from the Fe~{\sc{x}} ions, the derived magnetic field strengths should be the field strengths at heights of strong Fe~{\sc{x}} line emissions.
{However, the magnetic field strengths along the LOS are often inhomogeneous. Considering the optically thin nature of the coronal EUV emission, the derived magnetic field strengths should be the emission-weighted averaged field strengths.}
For the purpose of comparison, we defined the coronal magnetic field strength at each pixel in the model ($B_0$) as the emissivity-weighted field strength,
\begin{equation}
    B_0=\frac{\int_{LOS} \epsilon_{174}(s) \cdot B(s) ds}{\int_{LOS} \epsilon_{174}(s) ds}  \label{eq:em_b}
\end{equation}

\eqn{eq:em_b} can provide a precise estimation of the magnetic field strength when the field strength and emissivity are both constant within the line-formation height range along the LOS, or the emissions of the Fe~{\sc{x}} lines are from an infinitely thin layer along the LOS.
%
%
%
However, the Fe~{\sc{x}} line emissions in both the real corona and our model are from a wide range along the LOS. Nevertheless, \eqn{eq:em_b} still gives one of the best estimations of the coronal magnetic field strength in the model.
We took the vertical direction as the LOS and present the $B_0$ map in \figa{f9}(a), which shows that the coronal magnetic field strengths range from about one hundred to a thousand Gauss. Since the emission of an Fe~{\sc{x}} line roughly scales with the electron density squared, and the density generally decreases with height, $B_0$ should reflect the field strength in the lower corona.  

\subsection{Comparison between the coronal magnetic field strengths in the model and measured using the MIT technique} \label{subsec:b1}
Having determined the electron density and magnetic field strength under a fixed temperature using the technique introduced in \sect{subsec:diag_Trho_uni}, we can compare the measured magnetic field to the model field.
%
%
\figa{f9}~(b-f) presents the maps of magnetic field strengths obtained from the 257/174, 257/175, 257/177, 257/184, and 257/255 {\AA} line ratios, respectively.
The ratios involving the 174, 175, 177, and 184 {\AA} lines can only provide measurements of magnetic field at lower parts of the coronal loops, or around sunspot regions.
%
%
While for the 257/255 {\AA} line ratio, the MIT technique can yield a value of field strength at almost every pixel. However, the field strengths in the upper parts of loops are largely overestimated compared to the model values shown in \figa{f9}~(a).
%

\begin{figure} 
\centering {\includegraphics[width=80mm]{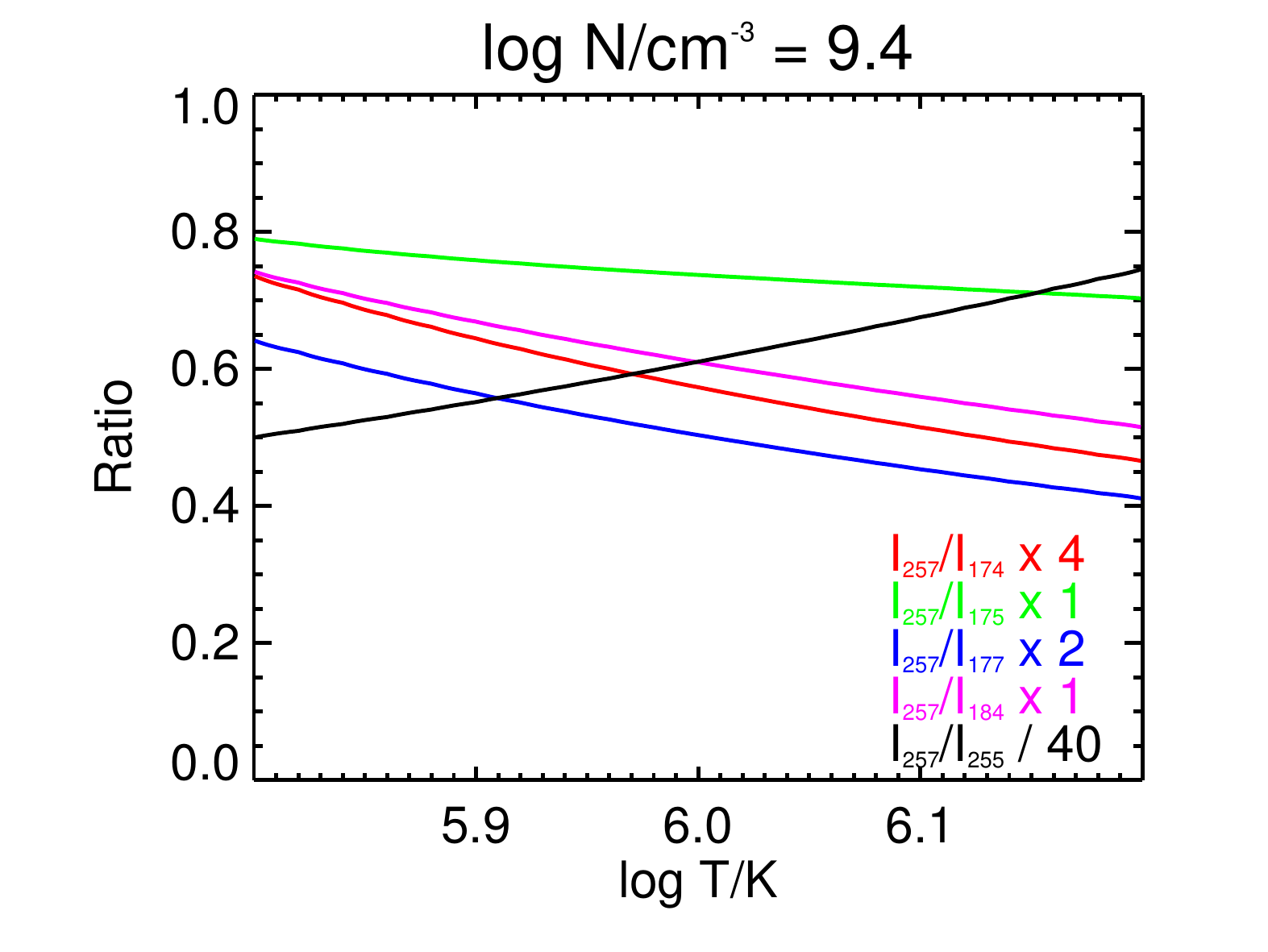}}
\caption{Intensity ratios of the Fe~{\sc{x}} 257/174, 257/175, 257/177, 257/184, 257/255 {\AA} line pairs as a function of temperature at a constant density of 10$^{9.4}$ cm$^{-3}$. 
The intensity ratios of different line pairs are scaled by different factors for a better illustration.
} \label{f12}
\end{figure}

\begin{figure*} 
\centering {\includegraphics[width=\textwidth]{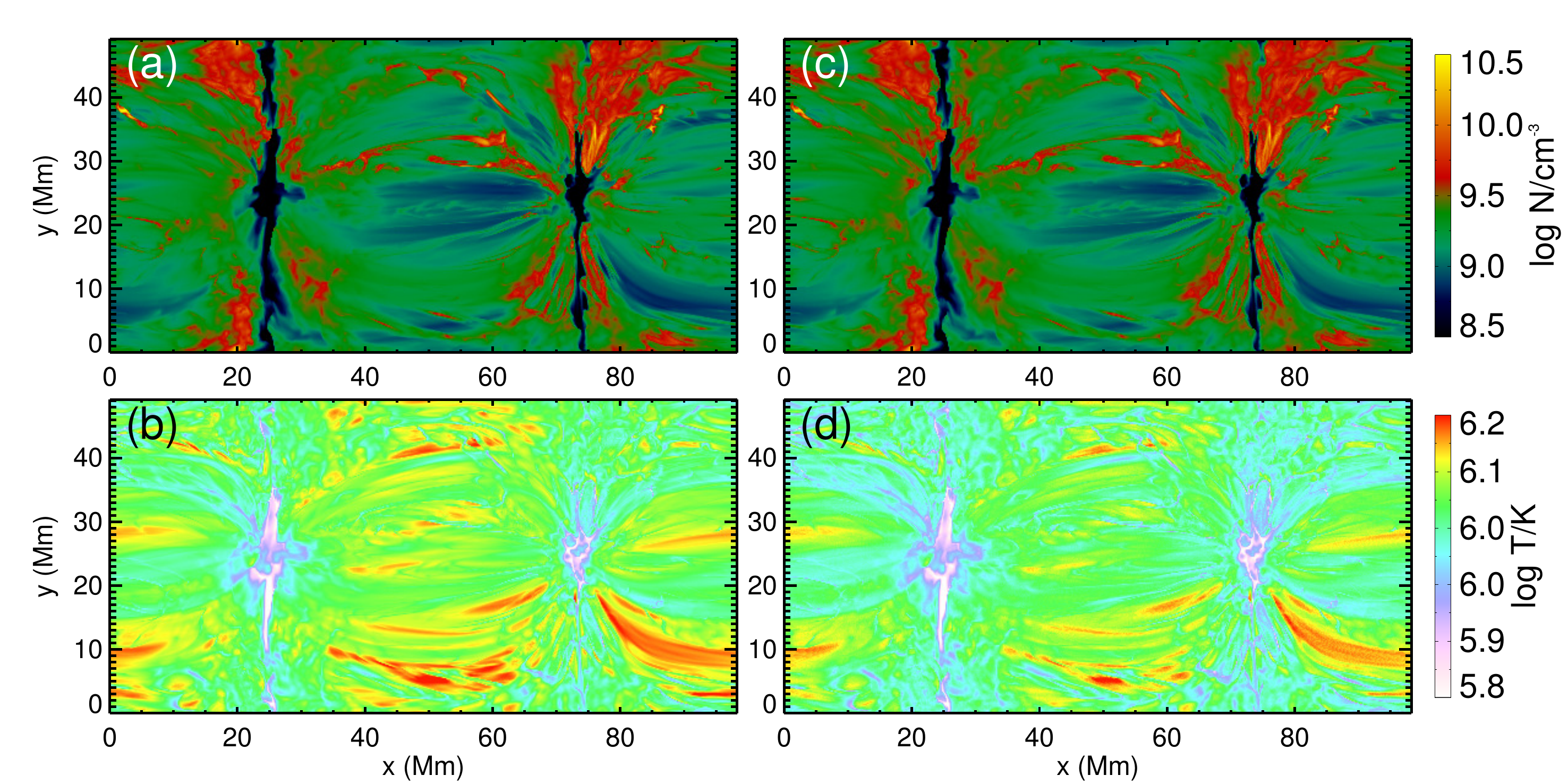}} 
\caption{(a--b) Maps of the emissivity-weighted density and temperature in the coronal model calculated using \eqns{eq:em_n} and (\ref{eq:em_T}), respectively. (c--d) Density and temperature maps obtained using the least--squares method introduced in \sect{subsec:diag_Trho_lsq}.
} \label{f4}
\end{figure*}

To explore the reasons why the MIT technique fails in regions far away from the sunspots and why the magnetic field measurements using the 255 {\AA} line and other reference lines behave differently, 
%
we present the intensity ratios of the 257/174, 257/175, 257/177, 257/184, and 257/255 {\AA} line pairs as a function of temperature in \figa{f12}.
We can see that all the line ratios are temperature-dependent, i.e., $I_{257}/I_{255}$ increases monotonically with temperature while the others decrease monotonically with temperature, meaning that the choice of temperature in the calculations will significantly affect the results of magnetic field measurements.
Furthermore, the temperature in the real coronal varies in space, and the formation temperature of an Fe~{\sc{x}} emission line at each pixel in the model can be different from the peak temperature of its contribution function \citep{Peter2006}.
To verify this, similar to \eqns{eq:em_h} and (\ref{eq:em_b}), we defined the emissivity-weighted coronal density and temperature in the model as:
\begin{equation}
    n_0=\frac{\int_{z} \epsilon_{174}(z) \cdot n_e(z) dz}{\int_{z} \epsilon_{174}(z) dz}  \label{eq:em_n}
\end{equation}
\begin{equation}
    T_0=\frac{\int_{z} \epsilon_{174}(z) \cdot T(z) dz}{\int_{z} \epsilon_{174}(z) dz}  \label{eq:em_T}
\end{equation}

The calculated density and temperature maps are presented in \figa{f4} (a) and (b), respectively. At each pixel these values could be regarded as the average density and temperature within the formation height range of the Fe~{\sc{x}} 174 {\AA} line. 
It is obvious that the emissivity-weighted temperature changes significantly at different locations.
As a result, if we underestimate the temperature at one location, the magnetic field strength obtained using the 257/255 {\AA} line ratio will be overestimated since $I_{257}/I_{255}$ increases monotonically with temperature (see \figa{f12}). While it will be underestimated using the other line ratios. Note that all these ratios increase monotonically with the magnetic field strength (e.g., \figa{f1}). 
%
Also, if the temperature is underestimated at locations of very weak magnetic field, we will fail to obtain a field strength value using the 257/174, 257/175, 257/177, and 257/184 {\AA} line ratios, as the ratio values will be lower than those corresponding to a field strength of zero Gauss (see \figa{f1}).
%
%
This exactly explains the behavior in the upper parts of the coronal loops, where the magnetic field and temperature are expected to be weaker and higher (\figa{f4}(b)), respectively.
Thus, the assumption of a fixed temperature we made in the calculations is not appropriate, 
and we have to determine the local temperature at each pixel for magnetic field diagnostics.

\section{MIT-based magnetic field measurements using the least--squares method for density and temperature determination} \label{sec:inver_lsq}

\begin{figure*} 
\centering {\includegraphics[width=14cm]{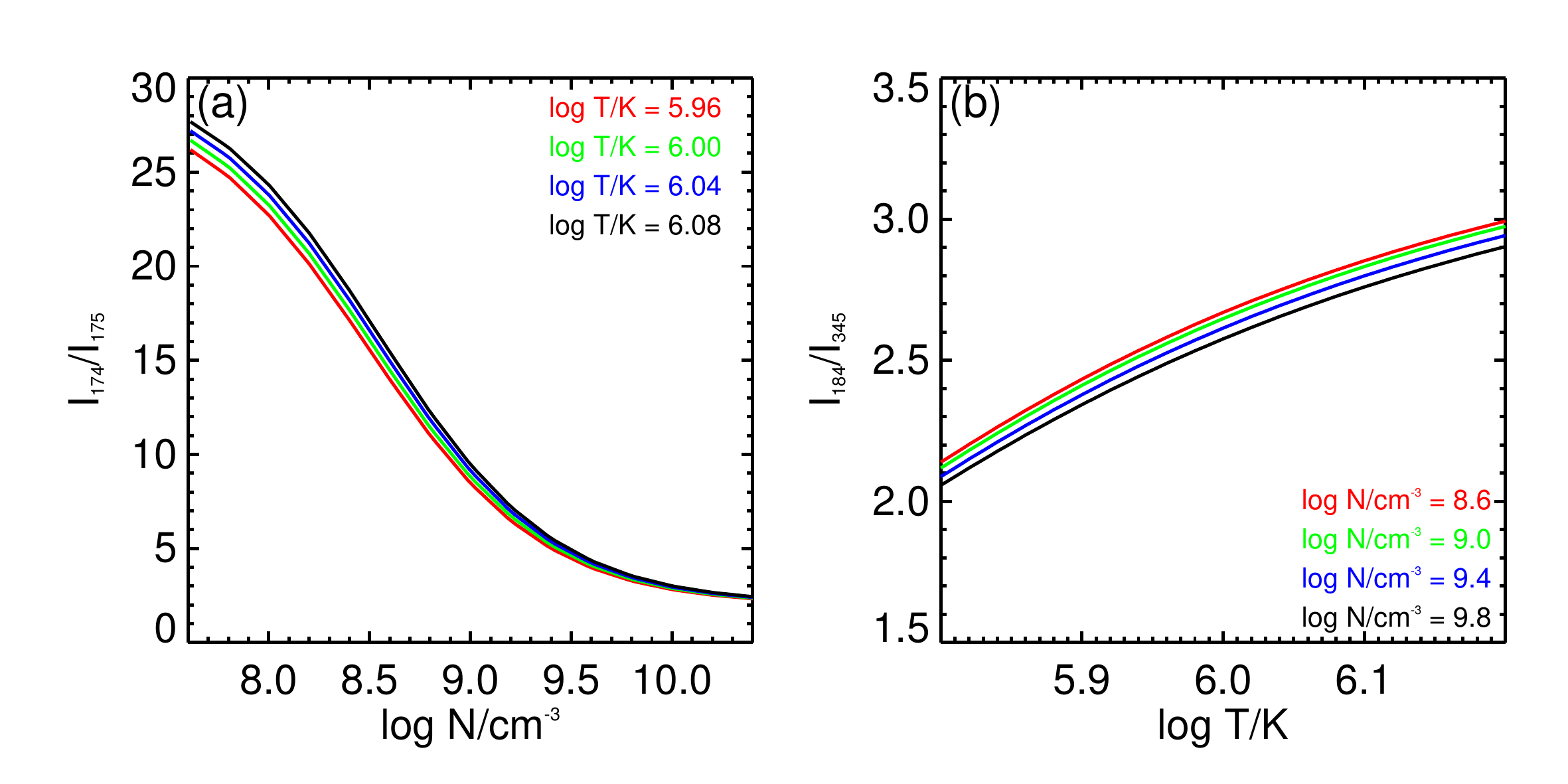}} 
\caption{(a) Intensity ratio of the 174 and 175 {\AA} lines as a function of electron density. Different curves correspond to results calculated at different temperatures. (b) Intensity ratio of the 184 and 345 {\AA} lines as a function of electron temperature. Different curves correspond to results calculated at different densities. 
} \label{f2}
\end{figure*}

\subsection{Density and temperature diagnostics} \label{subsec:diag_Trho_lsq}

As discussed in \sect{subsec:b1}, the uniform temperature distribution assumption has obvious limitations and introduces significant uncertainties in the magnetic field measurements. So, in addition to the density, we also need to determine the local formation temperature of the Fe~{\sc{x}} emission lines before the magnetic field measurement at each pixel.
%
%
%
%

The Fe~{\sc{x}} 174 or 184 {\AA} lines can provide temperature diagnostics when taken as a ratio with respect to any of the 257, 345, and 6374 {\AA} lines \citep{Del_Zanna2018}.
%
In this work, we adopted the 184/345 {\AA} line ratio for temperature measurements because
(1) the 257 {\AA} line involving the MIT transition is used to measure the magnetic field; 
(2) the 174 {\AA} line is used for density diagnostics; 
(3) the 6374 {\AA} line is difficult to be detected on the solar disk and {affected by photoexcitation. In addition}, the large wavelength difference between the 6374 {\AA} line and the 174 or 184 {\AA} line poses a formidable challenge on simultaneous well-calibrated observations.
%
%

\begin{figure*} 
\centering {\includegraphics[width=14cm]{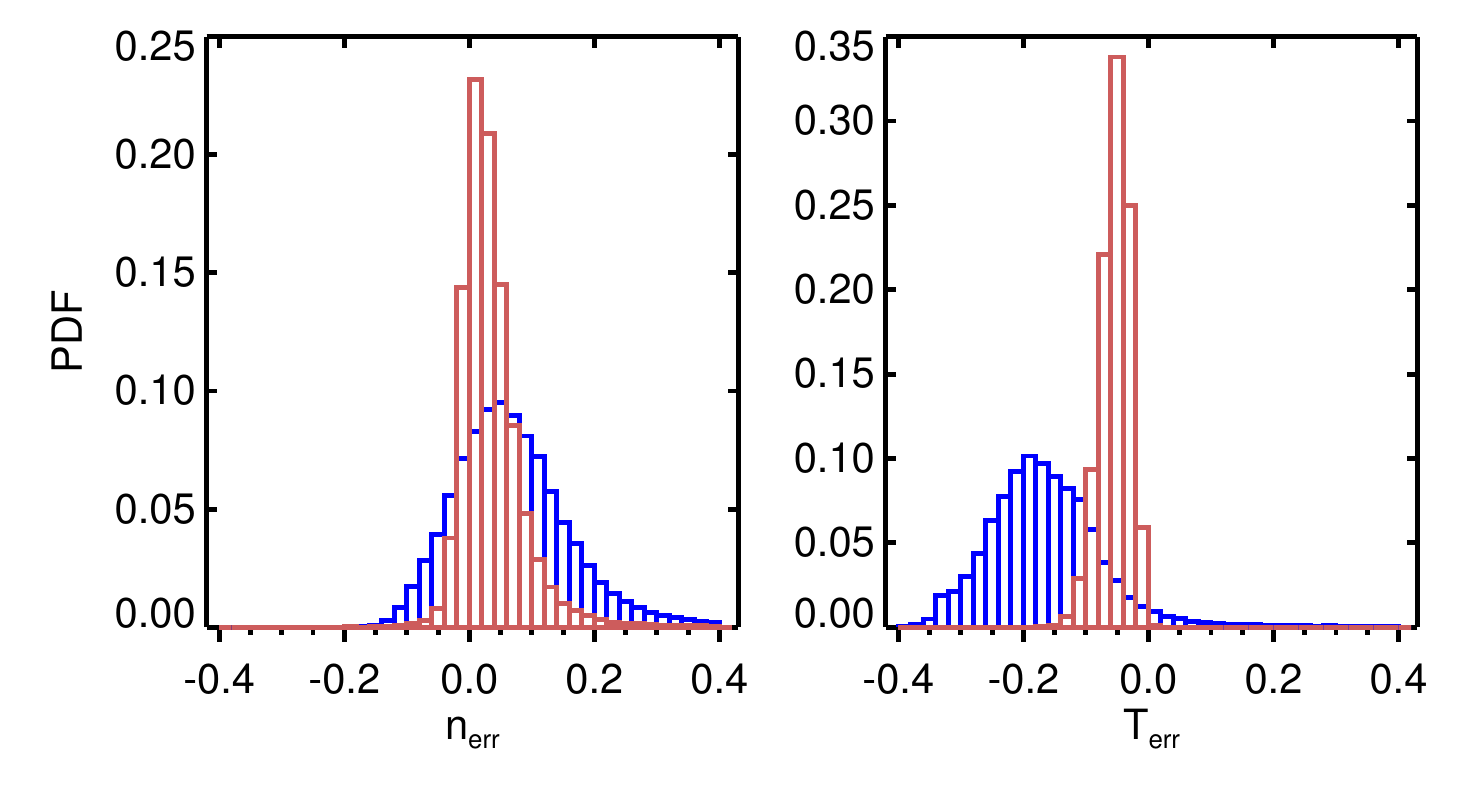}} 
\caption{Left: The histograms of $n_{err}$. The red and blue histograms are the results obtained from the least-squares method and assumption of a fixed temperature of 10$^6$ K, respectively.
Right: The same as the left panel but for $T_{err}$.
} \label{f16}
\end{figure*}

\figa{f2} presents the theoretical relationships between the 174 and 175 {\AA} ratio and electron density, and between the 184 and 345 {\AA} ratio and electron temperature. We can see that the 174/175 (184/345) {\AA} line ratio also changes slightly with the temperature (density).  %
%
To provide a simultaneous estimation of the density and temperature, we defined a norm value ($L$), based on the least-squares method, as
\begin{eqnarray}
    L=\left (\frac{G_{174}(T,n_e,B=0)}{G_{175}(T,n_e,B=0)}-\frac{I_{174}^*}{I_{175}^*}\right)^2\nonumber\\
    +\left(\frac{G_{184}(T,n_e,B=0)}{G_{345}(T,n_e,B=0)}-\frac{I_{184}^*}{I_{345}^*}\right)^2
    \label{r_ls}
\end{eqnarray}
where $I_{174}^*$, $I_{175}^*$, $I_{184}^*$, and $I_{345}^*$ are, respectively, intensities of the Fe~{\sc{x}} 174, 175, 184, and 345 {\AA} lines at each pixel.
%
%
%
The temperature ($T^*$) and density ($n_e^*$) can thus be simultaneously determined when $L$ reaches a minimum at each pixel. 
{To do this, we created a lookup table for $L$ at a large number of electron densities and temperatures and then selected the minimum value from the table.}
The maps of the determined density and temperature are shown in \figa{f4} (c-d). It is obvious that the derived density and thermal structures match very well with the ones calculated from \eqns{eq:em_n} and (\ref{eq:em_T}) (\figa{f4} (a-b)).
{Furthermore, we defined the differences between the measured temperature/density and the values obtained from \eqns{eq:em_n} and (\ref{eq:em_T}) as:
\begin{equation}
n_{err}=(n^*-n_0)/n_0
\end{equation}
\begin{equation}
T_{err}=(T^*-T_0)/T_0
\end{equation}
The histograms of $n_{err}$ and $T_{err}$ obtained using the least-squares method and the assumption of a fixed temperature of 10$^6$ K discussed in \sect{subsec:diag_Trho_uni} are shown in \figa{f16}.
We can see that the least--squares method can significantly reduce the differences. Besides, the differences in density and temperature are less than 10\% for more than 95\% pixels using the least--squares method,}
demonstrating that the least--squares method can provide a reliable estimation of the coronal density and temperature.
Using $T^*$ and $n_e^*$ at each pixel, we can then determine the magnetic field strength using the MIT technique, i.e., from the line ratios between the 257 {\AA} line and other reference lines as shown in \eqn{r_257_i}.

\subsection{Forward modeling for disk-center observations}\label{subsec:ondisk}

\begin{figure*} 
\centering {\includegraphics[width=\textwidth]{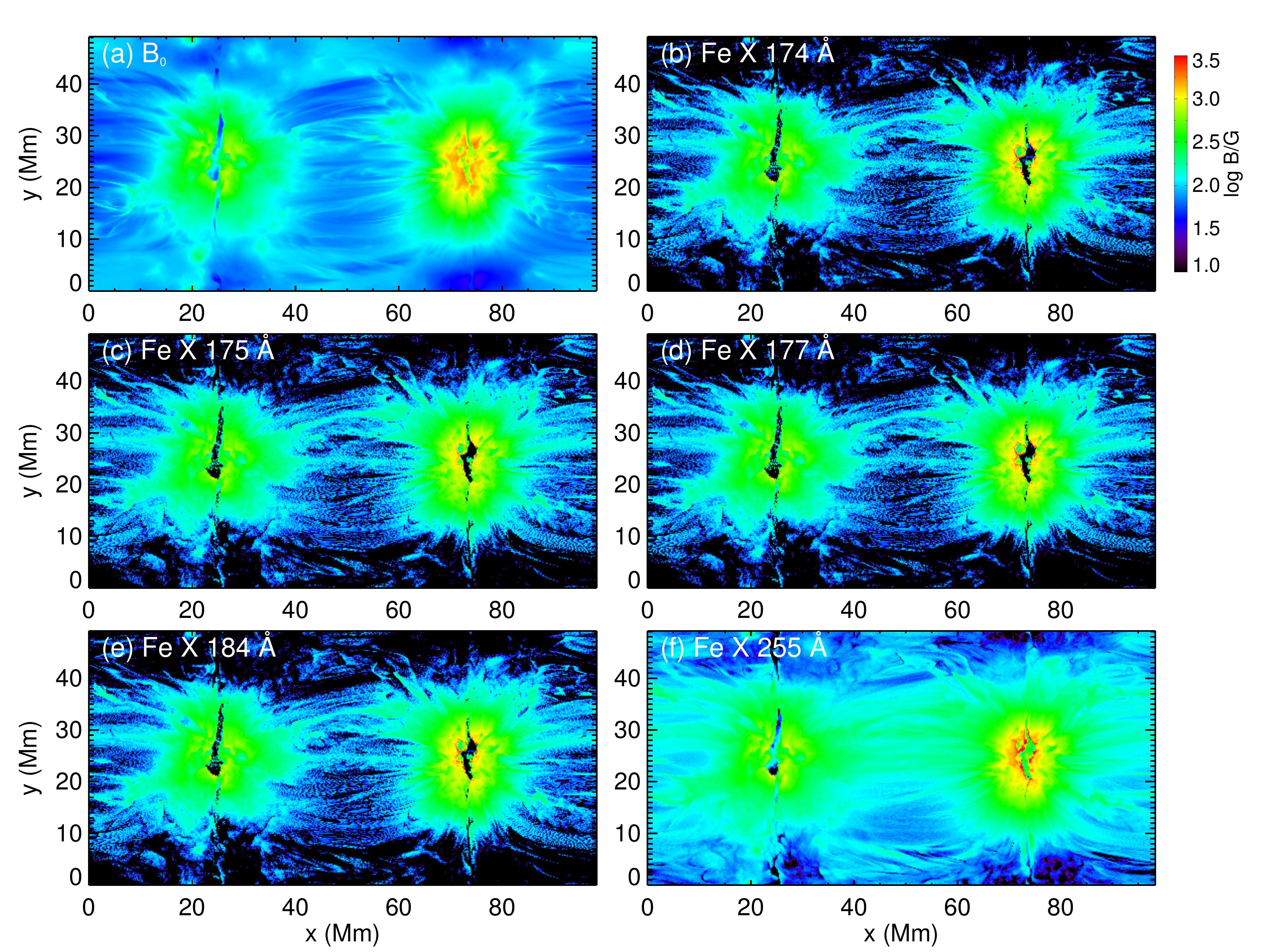}} 
\caption{
Similar to \figa{f9} but using the density and temperature maps shown in \figa{f4}(c--d). 
} \label{f5}
\end{figure*}

%
%
%
%
We first performed forward modeling assuming that the LOS is along the vertical ($z$) direction. The modeling result, shown in \figa{f5}, could help us evaluate the suitability of the MIT technique for measurements of the coronal magnetic field in disk-center observations.
It is obvious that the new method has significant improvement compared to the results in \figa{f9}. First, the new method provides a more accurate estimation of the magnetic field strength. Second, the new method can provide a measurement of the magnetic field strength not only around the footpoints but also in higher parts of the coronal loops.
%
%
However, we also noticed that similar to the results shown in \figa{f9}, the 257/174, 257/175, 257/177, and 257/184 {\AA} line ratios appear to underestimate the field strengths around the loop footpoints. While in the same regions the field strengths derived from the 257/255 {\AA} line pair are closer to the model values.
It is worth mentioning that our technique fails in regions without obvious Fe~{\sc{x}} emissions, and thus
the temperature, density, and magnetic field measurements show lanes at $x\approx 25$ and 75 Mm. We examined the magnetic field structures in these regions and found that the field lines are mostly vertical and extend to the top boundary of the simulation box. Continuous plasma outflows along these field lines evacuate materials in these regions, leading to very low densities and intensities in these regions. When the density is too low, the ${I_{257}}/{I_{i}}$ ratio becomes insensitive to the magnetic field (\figa{f1}(a)). In such a case, a reliable magnetic field measurement is impossible.  

\begin{figure*} 
\centering {\includegraphics[width=\textwidth]{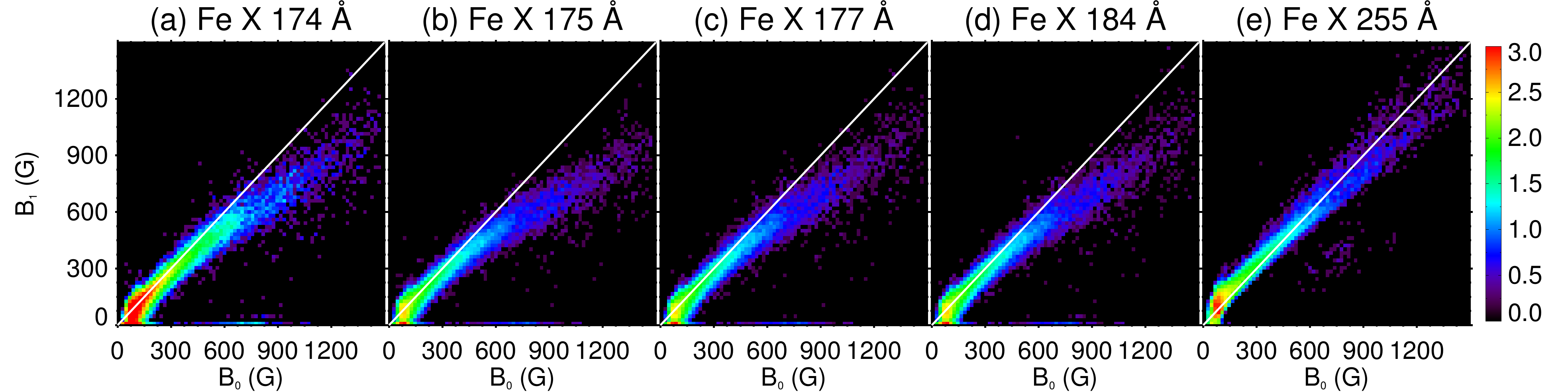}} 
\caption{Joint PDF of the coronal magnetic field strength in the MHD model ($B_0$) and the MIT-measured field strength ($B_1$). (a)-(e) correspond to results using the Fe~{\sc{x}} reference lines at 174, 175, 177, 184, and 255 {\AA}, respectively.
The white line represents the $B_0$ = $B_1$ line in each panel.
} \label{f6}
\end{figure*}

In order to understand the potential and limitations of this method better, we present the joint probability density functions (PDF) of the field strengths in the coronal model $B_0$ and the MIT-measured values $B_1$ in \figa{f6}.
%
The solid white line represents the $B_0=B_1$ reference line, i.e., the magnetic field strengths measured through the MIT technique are the same as the values in the model. In general, there is a good agreement between $B_0$ and $B_1$ for all the line pairs we used to diagnose the magnetic field.
We noticed that the magnetic field strengths are slightly underestimated when the field is stronger than $\sim$600 G.
This is because the density-sensitive 174/175 {\AA} line ratio is slightly dependent on the magnetic field strength, which was not considered in our calculation.
In the presence of an external magnetic field, the density derived using the above method was actually underestimated. From \figa{f2}(b) we can see that the temperature was also underestimated. For the 257/174, 257/175, 257/177, and 257/184 {\AA} line ratios, an underestimation of the density or temperature will lead to an underestimation of the magnetic field strength (\figa{f1}). This effect will be magnified at locations of strong magnetic field, i.e., around loop footpoints. While for the 257/255 {\AA} line ratio, an underestimation of the density and temperature will lead to an underestimation and overestimation of the field strength, respectively. In weak-field regions such as the top parts of coronal loops, the temperature effect dominates, leading to an overestimation of the field strength. Around the loop footpoints where the field is strong, the temperature and density effects may cancel out, which explains the fact that the field strengths derived from the 257/255 {\AA} line ratio are close to the model values there.

Nevertheless, \figa{f5} and \figa{f6} show that this technique can provide reasonably good measurements of the coronal magnetic field strengths in most parts of the active region. The measurements have a high degree of accuracy when the field strengths are in the range of a few hundred Gauss, which are the typical values of field strength in the lower corona of active regions. 

\begin{figure*}
\centering {\includegraphics[width=\textwidth]{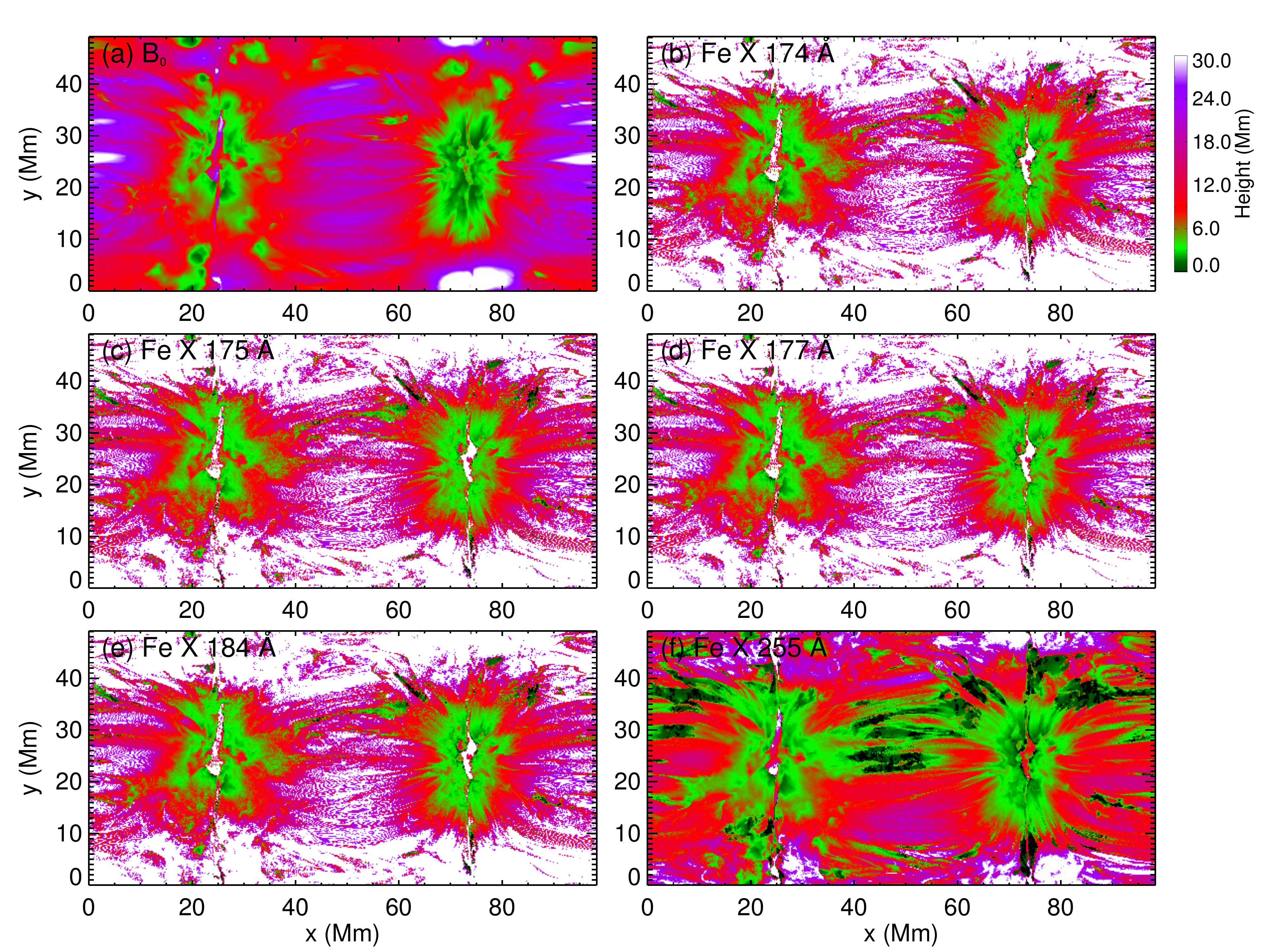}} 
\caption{The heights of the magnetic field shown in \figa{f5}.} \label{f13}
\end{figure*}


%
In addition, for each pixel we found the height $h_{B_0}$ where the model magnetic field is equal to the emissivity-weighted field strength $B_0$ calculated from \eqn{eq:em_b}.
For most pixels the magnetic field strength decreases with height, and we could easily found $h_{B_0}$. However, at some pixels there might be multiple heights where the model field is equal to $B_0$. In such cases, we took the largest height as $h_{B_0}$. The result is presented in \figa{f13}~(a). 
We found that $h_{B_0}$ increases from below 6 Mm near the footpoints of coronal loops to above 20 Mm in upper parts of the loops, which is roughly consistent with the average formation height of the Fe~{\sc{x}} 174 {\AA} line ($h_{em}$) shown in \figa{f3}~(c).
For the purpose of comparison, we also found the height ($h_{B_1}$) where the model magnetic field is equal to the MIT-measured field $B_1$ at each pixel (see \figa{f13} (b--f)).
For all the line pairs, $h_{B_1}$ distributions roughly match the $h_{B_0}$ distribution. Around the sunspots regions (footpoints and lower parts of the coronal loops), the heights are 3--9 Mm.
The heights of the magnetic field calculated using the 257/174, 257/175, 257/177, and 257/184 {\AA} line pairs are higher than 20 Mm around the loop tops, which are comparable to the values of $h_{B_0}$.
The 257/255 {\AA} line pair overestimate the magnetic field strengths at loop tops. Since the field strength decreases with height, the heights where the model field strengths equal to the MIT-measured field strengths are lower than $h_{B_0}$.

\subsection{Forward modeling for off-limb observations}\label{subsec:offlimb}

\begin{figure*} 
\centering {\includegraphics[width=12cm]{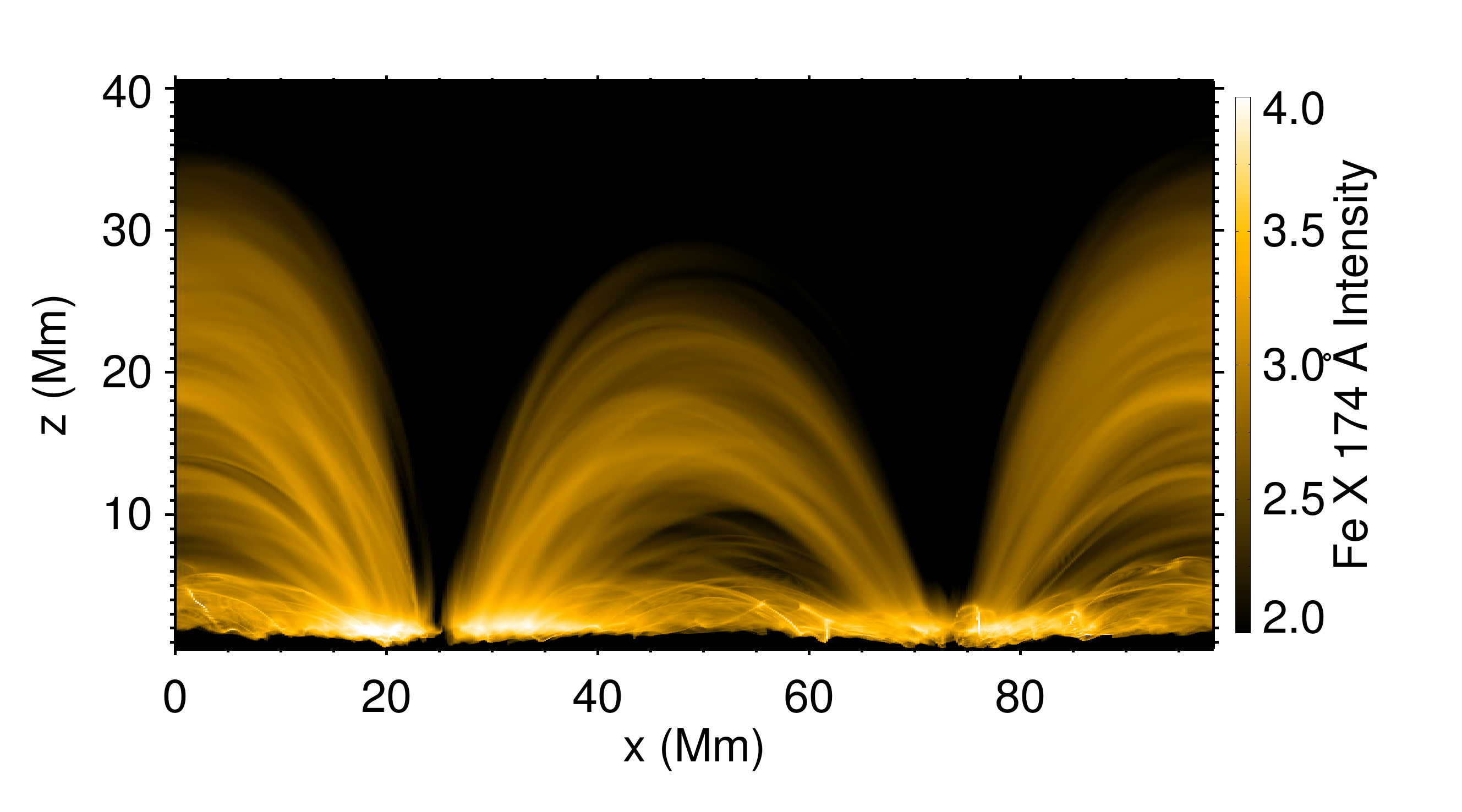}} 
\caption{The integrated emission of the Fe~{\sc{x}} 174 {\AA} line along the y-axis. The intensity is shown in logarithm scale and arbitrary unit.} \label{f15}
\end{figure*}

\begin{figure*} 
\centering {\includegraphics[width=\textwidth]{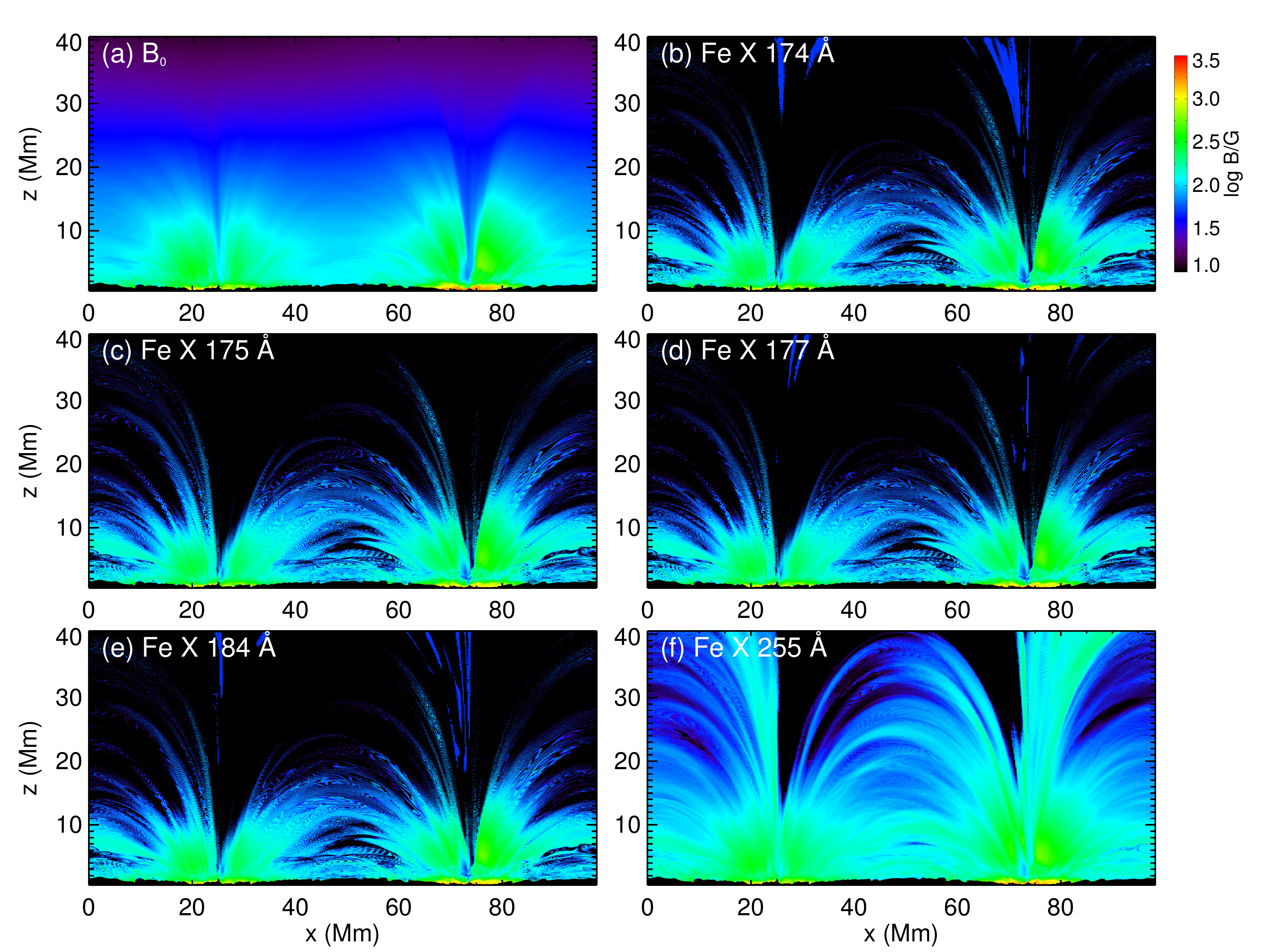}} 
\caption{Similar to \figa{f5} but for a LOS along the y-axis.} \label{f14}
\end{figure*}

So far, all the analyses have assumed that the active region in the model is located at the disk center, i.e., the LOS is along the vertical (\textit{z}) direction.
In this section, we chose a different LOS, i.e., along the \textit{y}-direction, which corresponds to an off-limb observation of the active region.
By integrating the emissivity along this new LOS, the intensities of the Fe~{\sc{x}} lines were calculated (\eqn{eq:int}). 
As an example, the intensity map of the 174 {\AA} line is shown in \figa{f15}. 
%
%
We can see loop structures that are similar to coronal loops in off-limb coronal observations of the real Sun.
The coronal loops are rooted in the sunspots and extend to heights of over 20 Mm. In the sunspot regions and above, the majority of the Fe~{\sc{x}} emission originates from a height lower than 5 Mm, around the loop footpoints. While between the two sunspots, a significant fraction of the Fe~{\sc{x}} emission comes from the upper parts of the coronal loops (above 10 Mm). These are consistent with the findings shown in \figa{f3}~(c) as well as in \figa{f13}. The intensity images of other Fe~{\sc{x}} lines exhibit a similar pattern.

The same technique described in \sect{subsec:diag_Trho_lsq} was applied to derive the temperature, density, and magnetic field strength. A comparison between the model magnetic field and the MIT-measured field is presented in \figa{f14}. \figa{f14}~(a) shows the $B_0$ map in the coronal model and \figa{f14}~(b-f) are the results obtained using the 257/174, 257/175, 257/177, 257/184, and 257/255 {\AA} line pairs, respectively.
%
%
Overall, the MIT technique provides a reasonably good estimation of the magnetic field in the coronal loop structures. 
Similar to the results in \sect{subsec:ondisk}, the MIT-measured magnetic field strengths in the lower parts of coronal loops agree well with the field strengths in the coronal model; the 257/255 {\AA} line pair appear to overestimate the magnetic field at the loop tops. At larger heights ($z\geqslant$ 20 Mm) and in regions directly above the sunspots ($x\approx 25$ and 75 Mm), the electron densities and Fe~{\sc{x}} line intensities are very low. From \figa{f1}(a) we can see that the ${I_{257}}/{I_{i}}$ ratio becomes insensitive to the magnetic field at very low densities. So in these regions our technique could not provide reliable measurements of the magnetic field.

\subsection{Selection of the Fe~{\sc{x}} lines}\label{subsec:select_linepairs}

In summary, the new technique developed in \sect{subsec:diag_Trho_lsq} was applied to
both synthetic disk-center (see \sect{subsec:ondisk}) and off-limb (see \sect{subsec:offlimb}) observations and provided an overall good estimation of the coronal magnetic field. 
However, there are some limitations caused by the choice of the Fe~{\sc{x}} lines when applying to real observations.
%

A critical point to note is that different lines used in this work are often observed by different detectors or even by different instruments. 
For example, among the line pairs 184/345 and 174/175 {\AA} which are used for the temperature and density measurements, the 345 {\AA} line was often not observed with the other three lines using the same instrument. This poses a formidable challenge on the relative radiometric calibration of different instruments.
%
%
%

%
The line pairs used to derive the magnetic field have been observed simultaneously by \textit{Hinode}/EIS but using different detectors.
The 257 and 255 {\AA} lines are observed by the long-wavelength detector, while the other reference lines, i.e., 174, 175, 177, and 184 {\AA} lines, are taken by the short-wavelength detector. 
This makes the relative calibration for the two detectors a critical factor that could affect precise measurements of the coronal magnetic field. Radiometric calibration has been proved to be one of the main sources of uncertainties in the measured magnetic field \citep{Landi2020b}.
In this respect, 257/255 {\AA} is the best choice for magnetic field measurements. However, the Fe~{\sc{x}} 255 {\AA} line is often too weak to have enough signal-to-noise ratios for spectral analysis.
%
%

The limitations of current EUV spectrometers impede the application of the least-squares diagnostic technique and thereby magnify the uncertainties of the magnetic field measurements based on the MIT method.
Thus, next-generation EUV spectrometers that can simultaneously observe the Fe~{\sc{x}} lines mentioned above and allow an accurate radiometric calibration in a wide range of wavelength, e.g., 170--350 {\AA}, are highly desired.

\section{Effect of the Fe~{\sc{x}} 3p$^4$ 3d $^4$D$_{5/2}$ and $^4$D$_{7/2}$ energy separation on magnetic field determination} \label{sec:dE}

\begin{figure} 
\centering {\includegraphics[width=80mm]{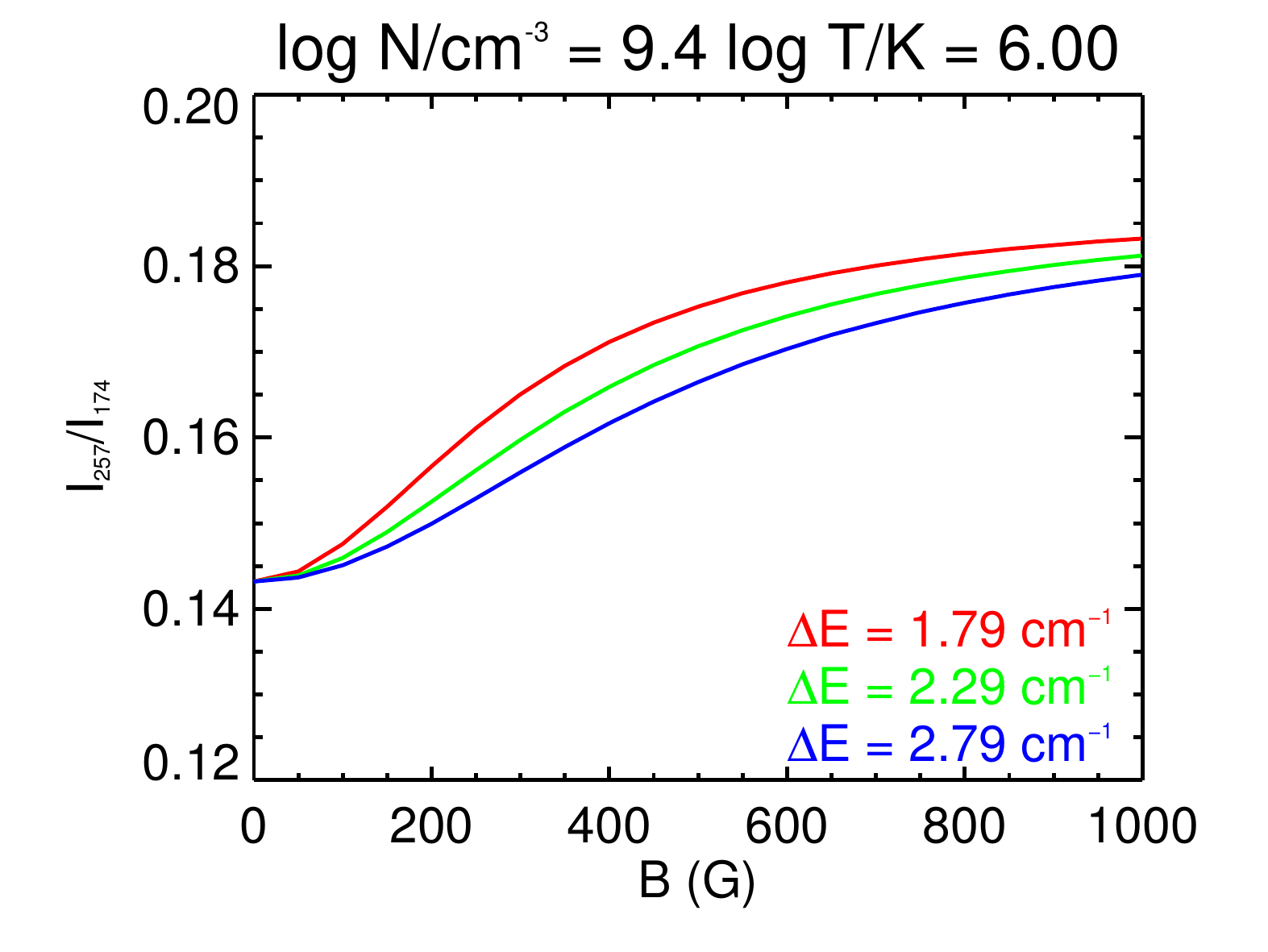}}
\caption{The intensity ratio of the Fe~{\sc{x}} 257 and 174 {\AA} lines as a function of magnetic field strength for different Fe~{\sc{x}} 3p$^4$ 3d $^4$D$_{5/2}$ and $^4$D$_{7/2}$ energy separations ($\Delta E$). The curves are plotted at a density of 10$^{9.4}$ cm$^{-3}$ and a temperature of 10$^{6.0}$ K. 
} \label{f11}
\end{figure}

The accuracy of the atomic database is critical for magnetic field measurements using the MIT diagnostic technique. One of the most critical factors is the uncertainty of the energy separation $\Delta E$ between the Fe~{\sc{x}} 3p$^4$ 3d $^4$D$_{5/2}$ and $^4$D$_{7/2}$ levels \citep{Li2016,Judge2016,Si2020,Landi2020b}. The value of $\Delta E$ is so small that its precise measurement is extremely challenging \citep[e.g., ][]{Judge2016,Landi2020b}. The best estimation of $\Delta E$ so far is probably given by \citet{Landi2020a} as $2.29 \pm 0.50$ cm$^{-1}$, with an upper and lower limit of 2.79 and {1.79} cm$^{-1}$, respectively. This value was recently adopted for measurements of the coronal magnetic field by \citet{Landi2020a,Landi2021} and \citet{Brooks2021}. All the analyses in \sects{sec:inver_uni} and \ref{sec:inver_lsq} are based on the atomic database calculated by taking $\Delta E =2.29$ cm$^{-1}$. In this section, we investigated the effect of different $\Delta E$ values on measurements of the magnetic field.
%

First, we created three groups of lookup tables of the contribution functions for the Fe~{\sc{x}} lines by taking $\Delta E$ = 1.79, 2.29, and 2.79 cm$^{-1}$, respectively.
Assuming a typical coronal temperature of 10$^{6.0}$ K and density of 10$^{9.4}$ cm$^{-3}$, we present the intensity ratio between the 257 and 174 lines as a function of magnetic field strength for three different energy separations in \figa{f11}.
%
%
%
%
\begin{figure*} 
\centering {\includegraphics[width=10cm]{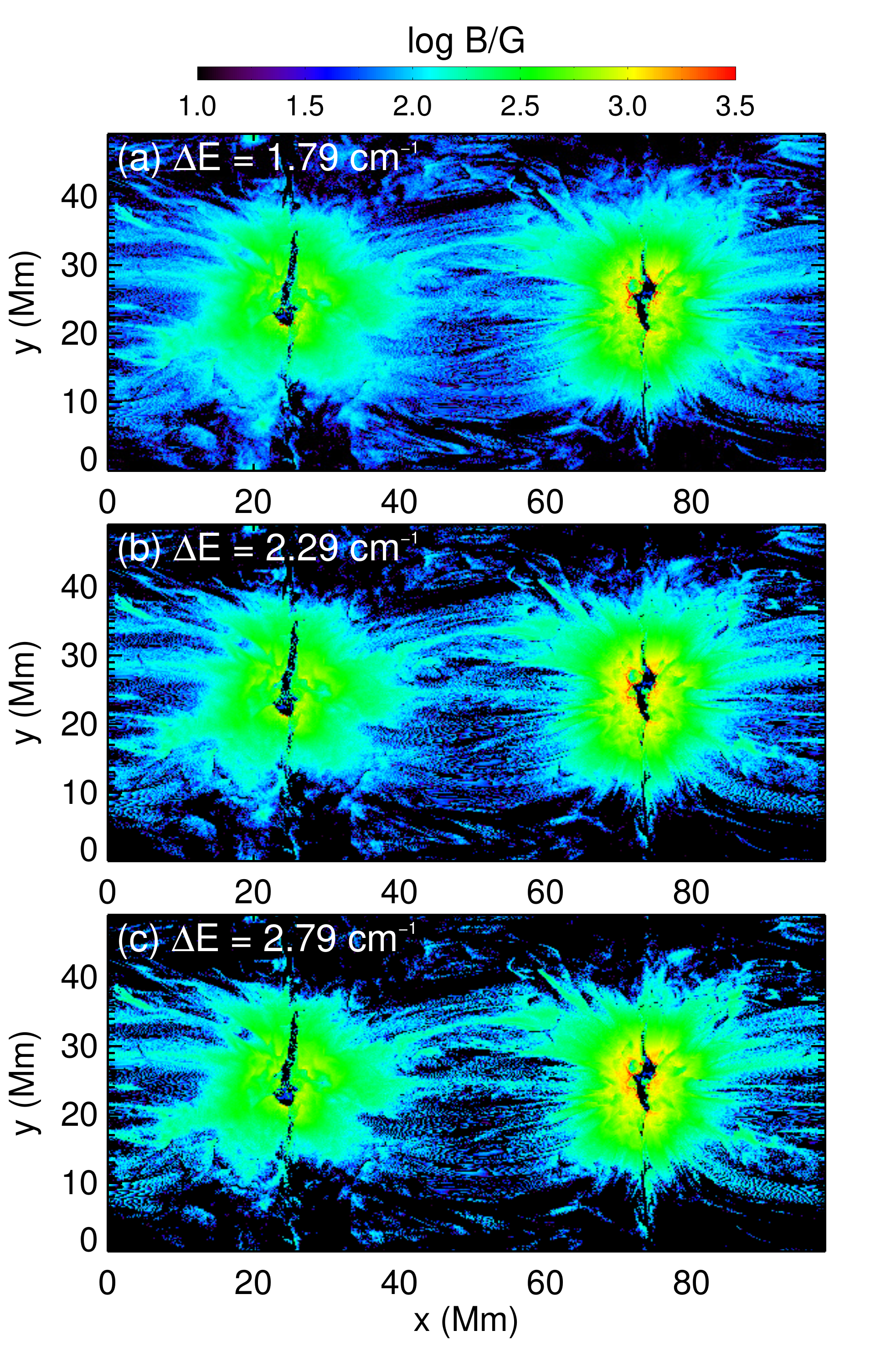}} 
\caption{Maps of magnetic field strength derived from the Fe~{\sc{x}} 257/174 {\AA} ratio for different energy separations $\Delta E$. 
(a) $\Delta E=$ 1.79 cm$^{-1}$; (b) $\Delta E=$ 2.29 cm$^{-1}$; (c) $\Delta E=$ 2.79 cm$^{-1}$.
Note that Panel (b) is identical to \figa{f5}(b).} \label{f7}
\end{figure*}
\begin{figure*} 
\centering {\includegraphics[width=\textwidth]{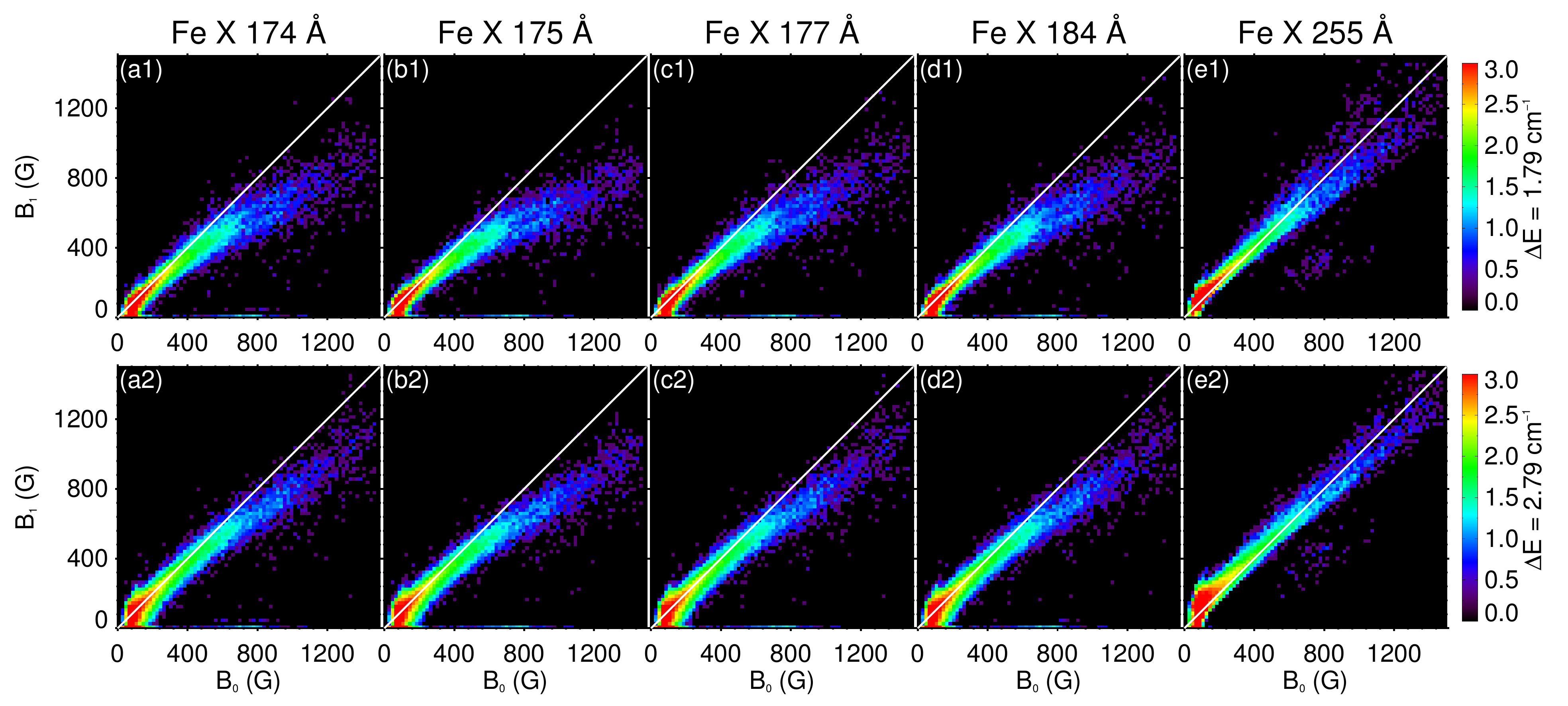}} 
\caption{Joint  PDF  of  the coronal  magnetic  field  strength  in  the  model  (B$_0$)  and  the MIT-measured field strength (B$_1$).
(a1--e1) Similar to \figa{f6} but for $\Delta E$ = 1.79 cm$^{-1}$. 
(a2--e2) Similar to (a1-e1) but for $\Delta E=$ 2.79 cm$^{-1}$.} \label{f8}
\end{figure*}
%
%
Then we applied the same technique employed in Section \ref{sec:inver_lsq} to derive the density, temperature and thereby the magnetic field with $\Delta E$ = 1.79, 2.29 and 2.79 cm$^{-1}$, respectively; the obtained magnetic field maps for a vertical LOS are presented in \figa{f7}.
Please note that the same $\Delta E$ value, thereby the same atomic database, was used for the Fe~{\sc{x}} line intensity synthesis and the magnetic field measurement.
Here we only present the results obtained from the 257/174 {\AA} line pair as an example, and the other line pairs give similar results.
We can see that the magnetic field maps shown in \figa{f7} are almost identical. 
Furthermore, we present the joint PDF of the coronal magnetic field in the MHD model ($B_0$ shown in \figa{f9}(a)) and the MIT-measured magnetic field ($B_1$) in \figa{f8} by taking $\Delta E$ = 1.79 (top panel) and 2.79 cm$^{-1}$ (lower panel), respectively. 
By comparing the results in \figa{f8} to \figa{f6} which corresponds to $\Delta E$ = 2.29 cm$^{-1}$, we found that the choice of $\Delta E$ does not significantly change the joint PDF of $B_0$ and $B_1$ derived from any specific line pair.
In other words, as long as we use the same $\Delta E$ value when measuring the magnetic field and synthesizing the Fe~{\sc{x}} emissions, the technique introduced in \sect{sec:inver_lsq} always provides a reasonable estimation of the coronal magnetic field regardless of the quantity of the energy separation.

\begin{figure*} 
\centering {\includegraphics[width=10cm]{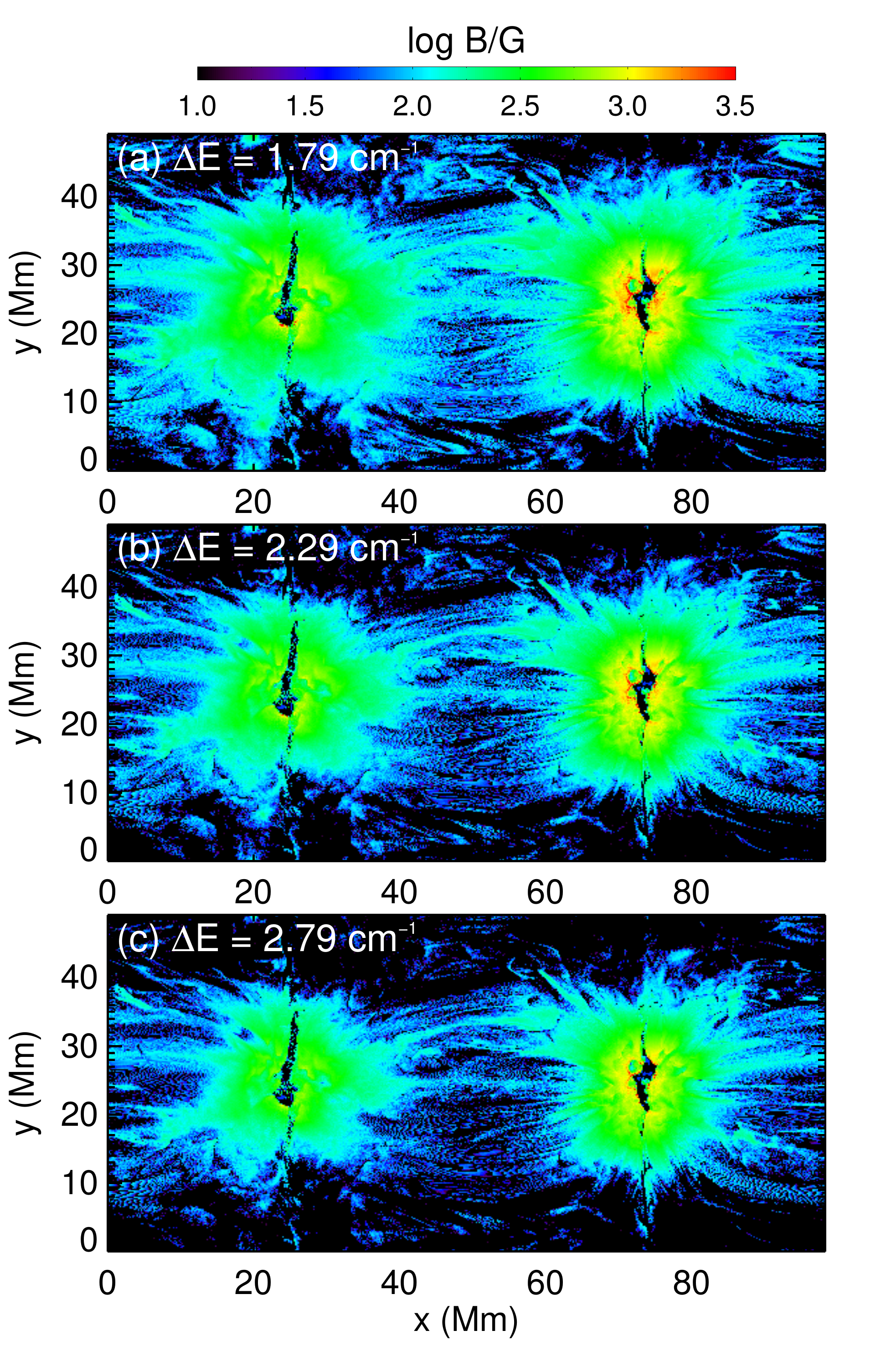}} 
\caption{Similar to \figa{f7} but taking $\Delta E=$ 1.79 (a), 2.29 (b), and 2.79 cm$^{-1}$ (c) when calculating the Fe~{\sc{x}} line intensities and taking $\Delta E=$ 2.29 cm$^{-1}$ when estimating the temperature, density, and magnetic field.
} \label{f7_v2}
\end{figure*}

\begin{figure*} 
\centering {\includegraphics[width=\textwidth]{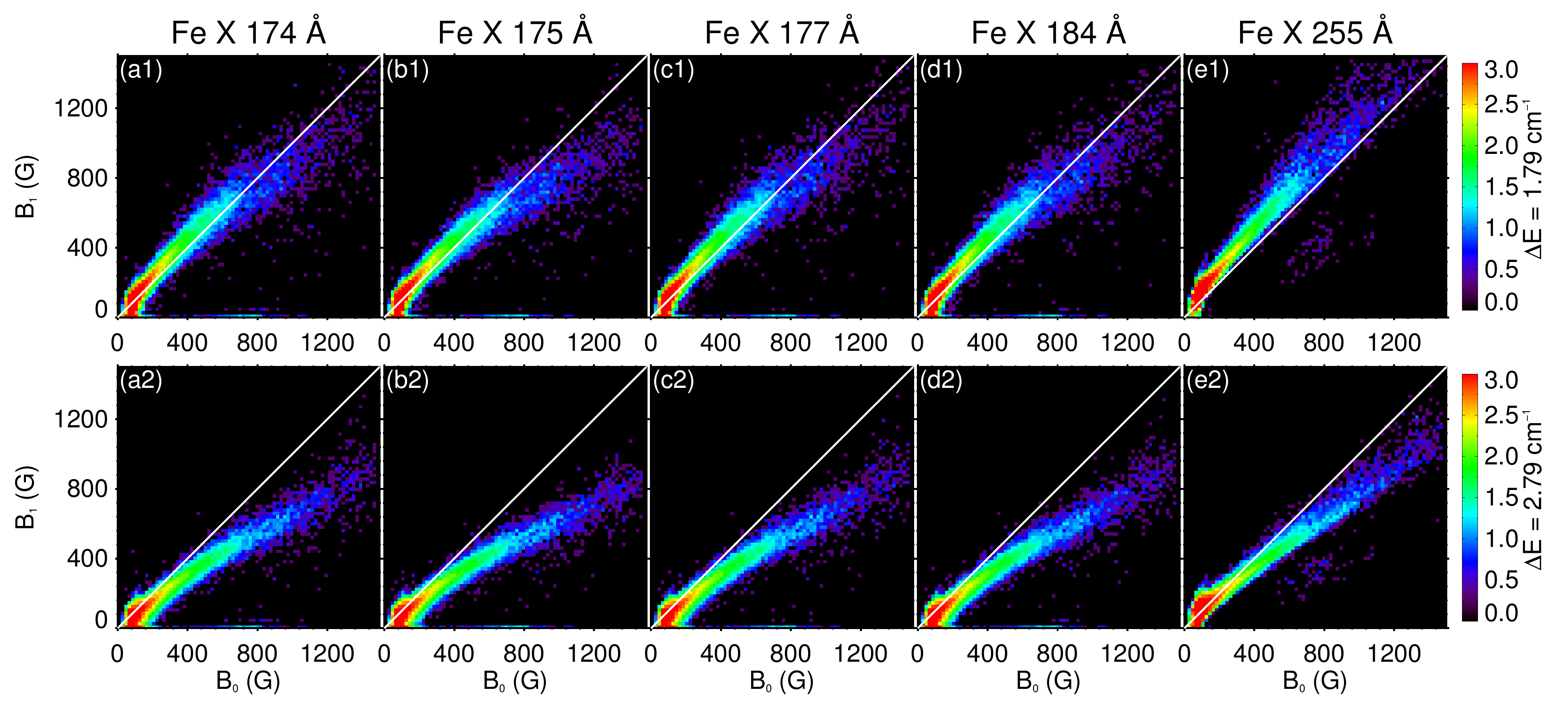}} 
\caption{Similar to \figa{f8} but taking dE $=$ 1.79 (a1--e1) and 2.79 cm$^{-1}$ (a2--e2) when calculating the Fe~{\sc{x}} line intensities and taking $\Delta E=$ 2.29 cm$^{-1}$ when estimating the temperature, density, and magnetic field.
} \label{f8_v2}
\end{figure*}

%
%
To further investigate the effect of the uncertainty in energy separation on the magnetic field diagnostics, we assumed a real $\Delta E$ value of 1.79, 2.29, and 2.79 cm$^{-1}$, respectively, for the line intensity synthesis, but instead adopted the commonly used $\Delta E=2.29$ cm$^{-1}$ for the density, temperature and magnetic field measurements.
%
%
The derived magnetic field maps using the 257/174 {\AA} line pair for different $\Delta E$ values are presented in \figa{f7_v2}, and the joint PDF are shown in \figa{f8_v2}.
Though a similar pattern is seen in the three magnetic field maps, noticeable differences in the magnitudes of the field strength clearly exist at some locations (\figa{f7_v2}). The different slopes that are present in the joint PDFs shown in \figa{f8_v2} and \figa{f6} suggest that an overestimation (underestimation) of the energy difference will result in an systematic overestimation (underestimation) of the coronal magnetic field. Thus, an accurate measurement of the energy difference between the $^4$D$_{5/2}$ and $^4$D$_{7/2}$ levels is essential for reducing the uncertainties in the magnetic field measurements.

\section{Summary and future perspectives} \label{sec:sum}

In this study, we have taken a 3D radiation MHD model of a solar active region and performed forward modeling to evaluate the accuracy of the magnetic field measurement based on the MIT technique.
We first synthesized the Fe~{\sc{x}} 174, 175, 177, 184, 255, 257, and 345 {\AA} line intensities from the modeled corona, and then employed two methods to estimate the electron density and temperature at each pixel. For the first method, we assumed a uniform temperature of 10$^{6.0}$ K and derived the density using the 174/175 {\AA} line ratio. For the second method, we determined the temperature and density from the 174/175 and 184/345 {\AA} line ratios based on the least--squares technique.
After estimating the density and temperature at each pixel, the magnetic field strength was determined using the intensity ratio between the Fe~{\sc{x}} 257 {\AA} line and another reference line based on the MIT theory.
We compared the MIT-measured magnetic field with the emissivity-weighted coronal magnetic field in the MHD model, and found that the least--squares method provides a much better estimation of the temperature, density, and thereby the magnetic field strength, suggesting that the simultaneous temperature determination is of critical importance for coronal magnetic field measurements using the MIT technique.
We found that all the investigated Fe~{\sc{x}} line pairs, i.e., 257/174, 257/175, 257/177, 257/184, and 257/255 {\AA}, can provide reasonably good estimations of the coronal magnetic field strength. 
We have also investigated the effect of the uncertainty in the Fe~{\sc{x}} 3p$^4$ 3d $^4$D$_{5/2}$ and $^4$D$_{7/2}$ energy separation $\Delta E$ on the magnetic field measurement, and found that choosing different $\Delta E$ values in the range of 1.79--2.79 cm$^{-1}$ could lead to systematic deviations from the coronal magnetic field in the MHD model.
To conclude, our investigation of forward modeling has verified that the MIT technique could provide reasonably accurate coronal magnetic field measurements in active region loops, demonstrating the potential of this technique in routine measurements of the coronal magnetic field.

It is important to note that this work is based on an ideal numerical experiment, in which some limitations of the real solar observations were not considered.
For instance, the Fe~{\sc{x}} lines adopted in our modeling have a wide range of wavelength, from 174 to 345 \AA. In real solar observations these spectral lines are often observed by different instruments or detectors. In such cases the uncertainty of radiometric calibration could lead to an additional uncertainty in the magnetic field measurements.
Also, we did not consider the insufficient signal-to-noise ratios and limited spectral resolutions that might be the case in real observations.
Finally, we did not examine how accurate the current atomic database is when applying it to the real Sun.
%
%

In order to achieve more reliable measurements of the coronal magnetic field using the MIT technique, more efforts need to be made in the future. Here we list several ideas that should be considered,
\begin{enumerate}
    \item Temperature determination is crucial for the coronal magnetic field measurement. 
    We used the Fe~{\sc{x}} 184/345 {\AA} line ratio to estimate the coronal temperature in this study.
    However, the 345 {\AA} line is not observed by \textit{Hinode}/EIS.
    Ideally, a spectrometer that can simultaneously observe the Fe~{\sc{x}} 174, 175, 184, 257, and 345 {\AA} lines would be needed.
    Alternatively, we may find other methods to determine the formation temperatures of the Fe~{\sc{x}} lines from available spectroscopic observations. 
    For example, {we may derive the formation temperature of the Fe~{\sc{x}} lines from the differential emission measure analysis. In addition,} we may take the temperature derived from the intensity ratio of the Fe~{\sc{ix}} 171.07/188.49 {\AA} or Fe~{\sc{xi}} 188.22/257.55 {\AA} lines that are formed at similar temperatures, or take their average.
    \item Radiometric calibration can cause significant uncertainties in the magnetic field diagnostics. The different Fe~{\sc{x}} lines are often observed by different detectors of Hinode/EIS. Improvements of the relative intensity calibration between different detectors are critical for the coronal magnetic field measurements.
    \item {Our analyses did not consider the noise and signal-to-noise ratio of the measured line intensities. These observational uncertainties may affect the applicability of the MIT technique, and thus should be evaluated in the future.}
    \item  The slight dependence of the density-sensitive line ratio 174/175 {\AA} on magnetic field should be properly accounted. The underestimation of field strengths around loop footpoints from the 257/174, 257/175, 257/177, and 257/184 {\AA} line pairs as well as the overestimation at loop tops from the 257/255 {\AA} line pair both appear to be caused by this effect.
    \item A precise knowledge of the energy separation between the Fe~{\sc{x}} 3p$^4$ 3d $^4$D$_{5/2}$ and $^4$D$_{7/2}$ levels $\Delta E$ is important for accurate magnetic field measurements.
    More analysis of spectral observations from SUMER and the Spectral Imaging of the Coronal Environment \citep[SPICE, ][]{SPICE} onboard Solar Orbiter may help to achieve a better accuracy in the determination of $\Delta E$.
    \item In addition to $\Delta E$, other atomic parameters such as the Einstein coefficients of all the energy levels of Fe~{\sc{x}} and collision excitation rates are also crucial \citep[e.g., ][]{Wang2020} for the synthesis of spectral intensities and diagnostics of plasma parameters. Advanced self-consistent large-scale calculations of the Fe~{\sc{x}} lines are challenging but necessary.
    \item The accuracy of the current atomic database needs to be tested by experiments in the laboratories. 
    Ongoing experiments using the electron beam ion trap \citep{Xiao2013} or similar instruments can help to verify the accuracy of the atomic data and provide constraints/corrections to the theoretical model.
\end{enumerate}

\begin{acknowledgments}
This work is supported by NSFC grants 11825301, 11790304, 12073004, 11704076 and U1732140, grant 1916321TS00103201, the Fundamental Research Funds for the Central Universities under grant 0201-14380041, and the CAS KLSA Open Research Program KLSA202117.
\end{acknowledgments}

\bibliography{mitsolar}{}
\bibliographystyle{aasjournal}

\end{document}